\documentclass[showpacs,twocolumn,aps,floatfix,
nofootinbib,superscriptaddress]{revtex4}

\usepackage{graphicx}
\usepackage{dcolumn}
\usepackage{amsmath}

\begin{document}
\title{Rotating Ay\'{o}n-Beato-Garc\'{i}a black hole as a particle accelerator}

\author{Sushant G. Ghosh}
\email{sghosh2@jmi.ac.in}
\affiliation{Centre for Theoretical Physics,
 Jamia Millia Islamia,  New Delhi 110025
 India}
\affiliation{Astrophysics and Cosmology Research Unit,
 School of Mathematics, Statistics and Computer Science,
 University of KwaZulu-Natal, Private Bag X54001,
 Durban 4000, South Africa}
 
\author{Pankaj Sheoran} 
\email{hukmipankaj@gmail.com}
\affiliation{Centre for Theoretical Physics,
 Jamia Millia Islamia,  New Delhi 110025
 India}
 
\author{Muhammed Amir}
\email{amirctp12@gmail.com}
\affiliation{Centre for Theoretical Physics,
 Jamia Millia Islamia,  New Delhi 110025
 India}

\begin{abstract}
We study the collision of two particles with equal masses moving in
the equatorial plane near horizon of the rotating regular Ay\'{o}n-Beato-Garc\'{i}a 
(ABG) black hole (BH) and calculate the center-of-mass (CM) energy for the colliding particles
for both extremal and non-extremal cases. It turns out that CM energy depends not only on 
rotation parameter $a$ but also on charge $Q$. Particularly for the extremal rotating regular 
ABG BH, CM energy of two colliding particles could be arbitrarily high for critical angular 
momentum of particles. Furthermore, we also show that, for a non-extremal BH, there exist 
a finite upper limit of CM energy, which changes with charge $Q$. A  comparison, with Kerr and 
Kerr-Newman black holes, is included.
\end{abstract}

\maketitle

\section{Introduction}
Recently, Ban$\tilde{a}$dos-Silk-West (BSW) \cite{Banados:2009pr} studied collision 
of two particles (e.g. dark matter particles) in the vicinity of the horizon 
of the Kerr black hole (BH) and found that center-of-mass (CM) energy of the  
colliding particles in the equatorial plane can be arbitrarily high in the limiting 
case of extremal BH.  This imply that the extremal rotating BH may be considered as a
Planck energy scale particle accelerator, which might allow us to explore ultra high energy collisions 
and astrophysical phenomena, such as the gamma ray bursts and the active galactic nuclei. 
In the work of BSW \cite{Banados:2009pr}, two particles of equal mass and equal energy 
falling freely from rest at infinity and approaching the extremal Kerr BH on the equatorial 
plane were considered. The energy in the CM frame was computed and the critical value of the 
particle angular momentum at which the energy blows up were determined. However, Jacobson and 
Sotiriou \cite{Jacobson:2009zg} elucidate the mechanism for this result, and  point out its 
practical limitations given that extremal BHs do not exist in nature, e.g., the spin $a$ of 
astrophysical BHs should not exceed $a/M= 0.998$ \cite{thorne} ($ M $ is the mass of the BH). 
In any case, the “third law” of BH thermodynamics (Bardeen et al. 1973) asserts that a BH cannot 
be spun-up in a finite time to the extreme spin value $a/M = 1$.  In particular, Jacobson and 
Sotiriou \cite{Jacobson:2009zg} demonstrated that infinite collision energy can only be attained 
at the horizon, and with a maximally spinning BH. Lake \cite{Lake:2010bq} also demonstrated that the CM 
energy of colliding particle near the inner horizon of a non-extremal Kerr BH is finite. It is known that 
the motion of a particle traveling in the background of a charged spinning BH depends not only on 
the spin but also on the charge of the BH. Therefore, the CM energy of collision expected to depend 
both on spin and charge.  The BSW scenario is generalized to charged BHs \cite{Zaslavskii:2010aw}, 
the Kerr-Newman family of BHs \cite{Wei:2010vca} and general rotating BHs \cite{Igata:2012js}. It is 
shown that, for a near-extremal charged BH, there always exists a finite upper bound for CM energy, 
which decreases with the increase of the charge $ Q $. 

The BSW effect near the event horizon of the Kerr-Taub-NUT BH was investigated by \cite{Liu:2010ja}, 
around the Kaluza-Klein (extremal) BH in \cite{Mao:2010di} and found the infinitely 
large CM energy near the horizon in both rotating and non-rotating cases. The BSW effect for the 
Einstein-Maxwell-dilaton-axion BH \cite{Hussain:2012zza}, and the BTZ BH was discussed in \cite{Yang:2012we}. 
In \cite{Zhu:2011ae}, the CM energy was generalized for charged particles moving in an electromagnetic field 
and braneworld BHs were discussed. Even for nonmaximally rotating BHs,  Grib and Pavlov 
\cite{Grib:2010dz,Grib:2010xj,Grib:2011ph,Grib:2013vc} argued that the CM energy for two particles 
collision can be unlimited for the non-maximal rotation as well when one considers the multiple scattering. 
A general explanation for the arbitrarily high CM energy is presented in terms of the Killing vectors 
and the Killing horizon by Zaslavskii \cite{Zaslavskii:2010jd}. Further, the author 
\cite{Zaslavskii:2012fh,Zaslavskii:2010pw,Zaslavskii:2012qy} clarified that the universal property 
of acceleration of particles near rotating BHs and give a general argument of this BSW 
mechanism for the general rotating BHs. Further, Joshi and Patil \cite{Patil:2011ya} found that 
the CM energy to be high for the naked singularity of the Kerr BH and other BHs \cite{Patil:2011uf}. 

More recently, BSW  mechanism is used to calculate the CM energy of collision for two  particles 
freely falling,  from rest at infinity, in the horizon of a Ay\'{o}n-Beato-Garc\'{i}a (ABG) BH 
\cite{Pradhan:2014oaa}.  It turns out that the CM energy for ABG BH can be infinitely high for 
the extremal case \cite{Pradhan:2014oaa}.  The rotating counterpart of ABG BH is another interesting 
and important spacetime \cite{Toshmatov:2014nya}, which is a solution of Einstein equations coupled 
to nonlinear electrodynamics. Besides the mass M and the rotation parameter $a$, the rotating ABG 
spacetime carries with the charge, $Q$.  It is widely believed that spacetime singularities do not 
exist in Nature, but that they represent a limitation of the classical theory. While we do not yet 
have any solid theory of quantum gravity models of BH solutions without singularities have been 
proposed \cite{bardeen,AyonBeato:1998ub,Hayward:2005gi}. These spacetimes have an event horizon and 
no pathological features like singularities or regions with closed timelike curves.  The rotating ABG 
metrics are more important as they can be tested by astrophysical observations, as the BH spin plays a 
fundamental role in any astrophysical process \cite{Bambi:2013ufa}.  The rotating ABG black hole are axisymmetric, 
asymptotically flat and depend on the mass and  spin of the black hole as well as on a parameter $Q$ that measure 
potential deviations from the Kerr metric. The rotating ABG metric includes the Kerr metric as the special case 
if this deviation parameter vanishes. 

The main purpose of this paper is to study the collision of two particles in the background of the rotating 
ABG spacetime and to see what the effects of the charge $Q$ on the CM energy for the particles in the near-horizon 
collision.  It turns out that our results can be reduced to the ones of BSW \cite{Banados:2009pr} as the charge parameter
turns to zero and nonrotating ABG BH when the rotation parameter $a=0$. Besides, we may be more interest to discuss 
CM energy of the rotating ABG BH because it does not have a simple horizon structure as the previous studied BHs. 

\section{Rotating Ay\'{o}n-Beato-Garc\'{i}a black hole}
In this section, we would like to study the rotating ABG BHs. The gravitational field of rotating ABG 
spacetime ~\cite{Toshmatov:2014nya} is described by metric which in the Boyer-Lindquist coordinates with units $c=G=1$ is given by
\begin{eqnarray}\label{abg}
d{s}^2 &=&-f(r,\theta) dt^2+\frac{\Sigma}{\Delta} dr^2
\nonumber \\ &&
-2a\sin^2\theta(1-f(r,\theta))d\phi dt+\Sigma d\theta^2 \nonumber \\ && 
+ \sin^2\theta[\Sigma-a^2(f(r,\theta)-2)\sin^2\theta]d\phi^2,
\end{eqnarray}
where the function $f(r,\theta)$ and $\Delta$ are given by
\begin{eqnarray}\label{fr}
f(r,\theta) &=& 1-\frac{2M r \sqrt{\Sigma}}{(\Sigma+Q^2)^{3/2}}+\frac{Q^2\Sigma}{(\Sigma+Q^2)^2},
\nonumber \\ 
\Delta &=& \Sigma f(r,\theta) + a^2\sin^2\theta
\end{eqnarray}
and $\Sigma=r^2+a^2\cos^2\theta$. The parameters $a$, $M$ and $Q$ are respectively correspond to 
rotation, mass and the electric charge. We shall demonstrate that, for certain range of values of 
$M$ and $Q$, the metric (\ref{abg}) is a BH. The curvature invariant $R$, $R_{ab}R^{ab}$ and
$R_{abcd}R^{abcd}$ for the metric (\ref{abg}) reveals that the rotating ABG BH is regular 
everywhere for $ a,M,Q \neq 0 $. The metric (\ref{abg}) is a non-singular rotating charged 
BH which generalizes the standard Kerr BH. In addition, if $Q=0$ the metric (\ref{abg}) reduce 
to Kerr BH \cite{kerr}. Further when both $a, Q = 0$ the metric (\ref{abg}) is Schwarzchild BH 
\cite{schw}. If $a\rightarrow 0$ we have an exact ABG BH \cite{Eloy} which is a solution of 
Einstein field equations coupled to the nonlinear electrodynamics. We know that the Kerr BH has 
two horizons like surfaces: static limit surface and event horizon. We are interested in these 
two surfaces for the rotating ABG spacetime metric (\ref{abg}). We start with calculating location 
and structure of static limit surface which requires the prefactor of $d{t}^2$ to vanish. Then it 
follows from (\ref{abg}) that the static limit surface will satisfy
\begin{eqnarray}\label{sls}
f(r,\theta)= (\Sigma+Q^2)^2-2Mr\sqrt{\Sigma(\Sigma+Q^2)}+{Q^2\Sigma} = 0.
\end{eqnarray}
The Eq.~(\ref{sls}) for $Q=0$ becomes exactly same as Kerr \cite{kerr}. The surface of no return 
is known as the event horizon. Beyond event horizon the gravitational force becomes so high that 
it is impossible for any object, even light to escape from the gravity of the BH. The spacetime 
(\ref{abg}) generically must have two horizons, viz., the inner or Cauchy horizon and the outer or
event horizon.   The horizons of the BH (\ref{abg}) are calculated by equating the $g^{rr}$ component 
of the metric (\ref{abg}) to zero, i.e.,
\begin{figure*}
\begin{tabular}{c c c}
\includegraphics[scale=0.65]{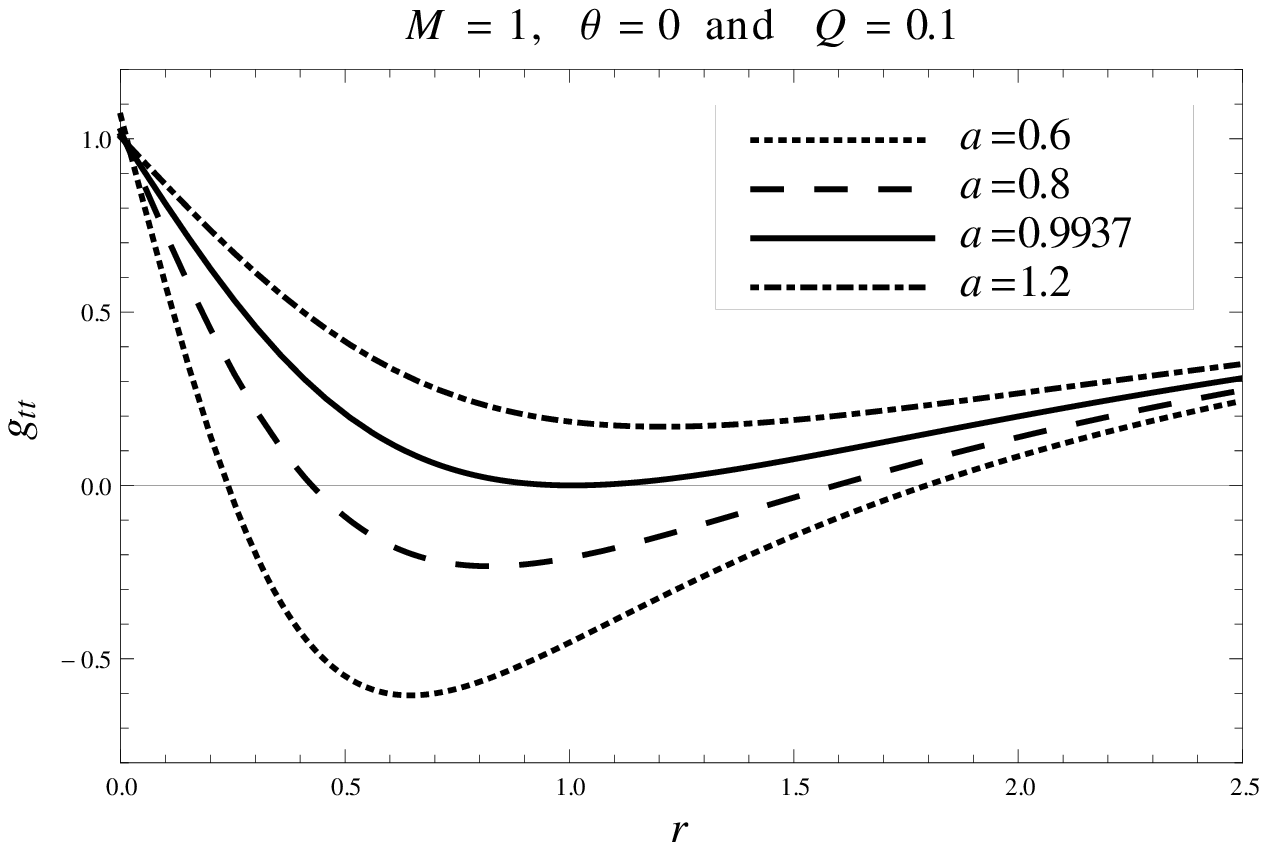}\hspace{-1cm}
&\includegraphics[scale=0.65]{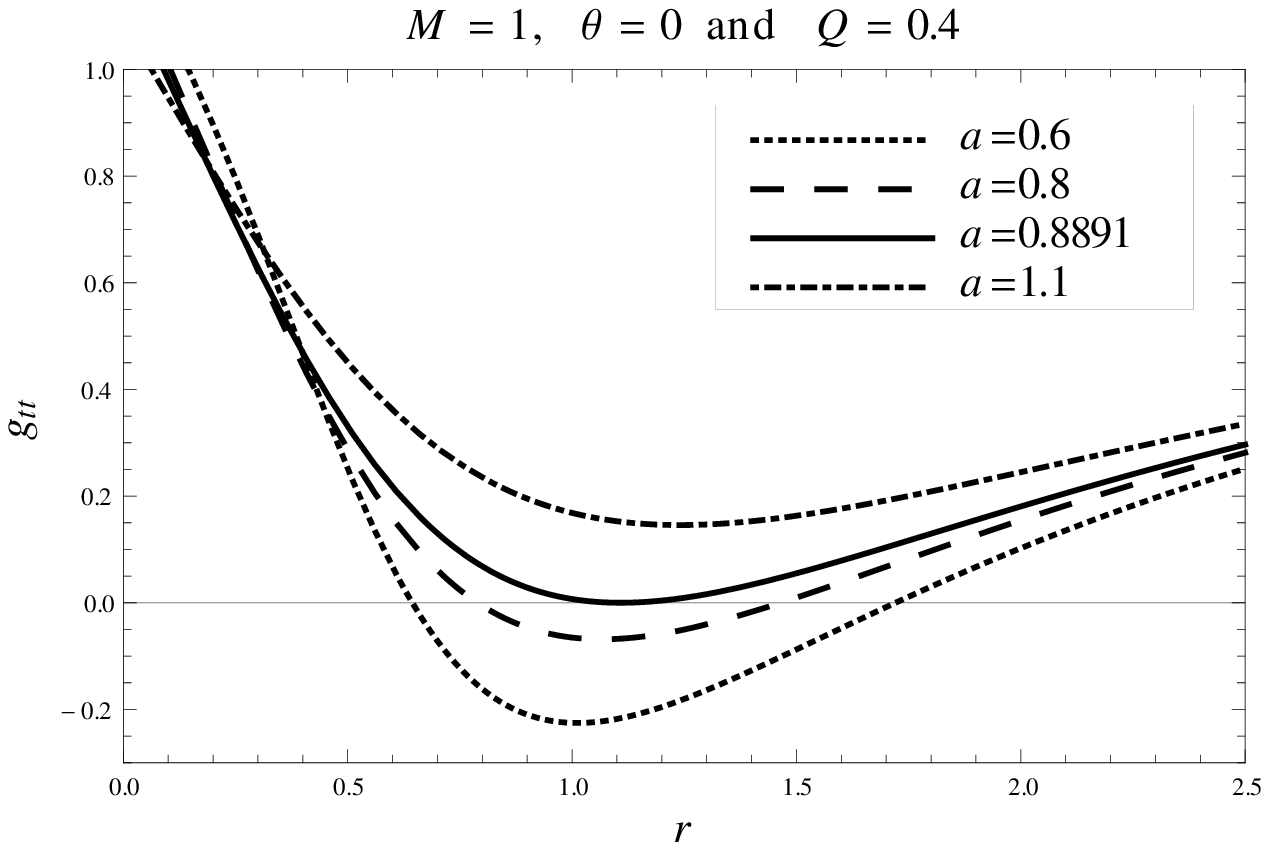}
 \\
 \includegraphics[scale=0.65]{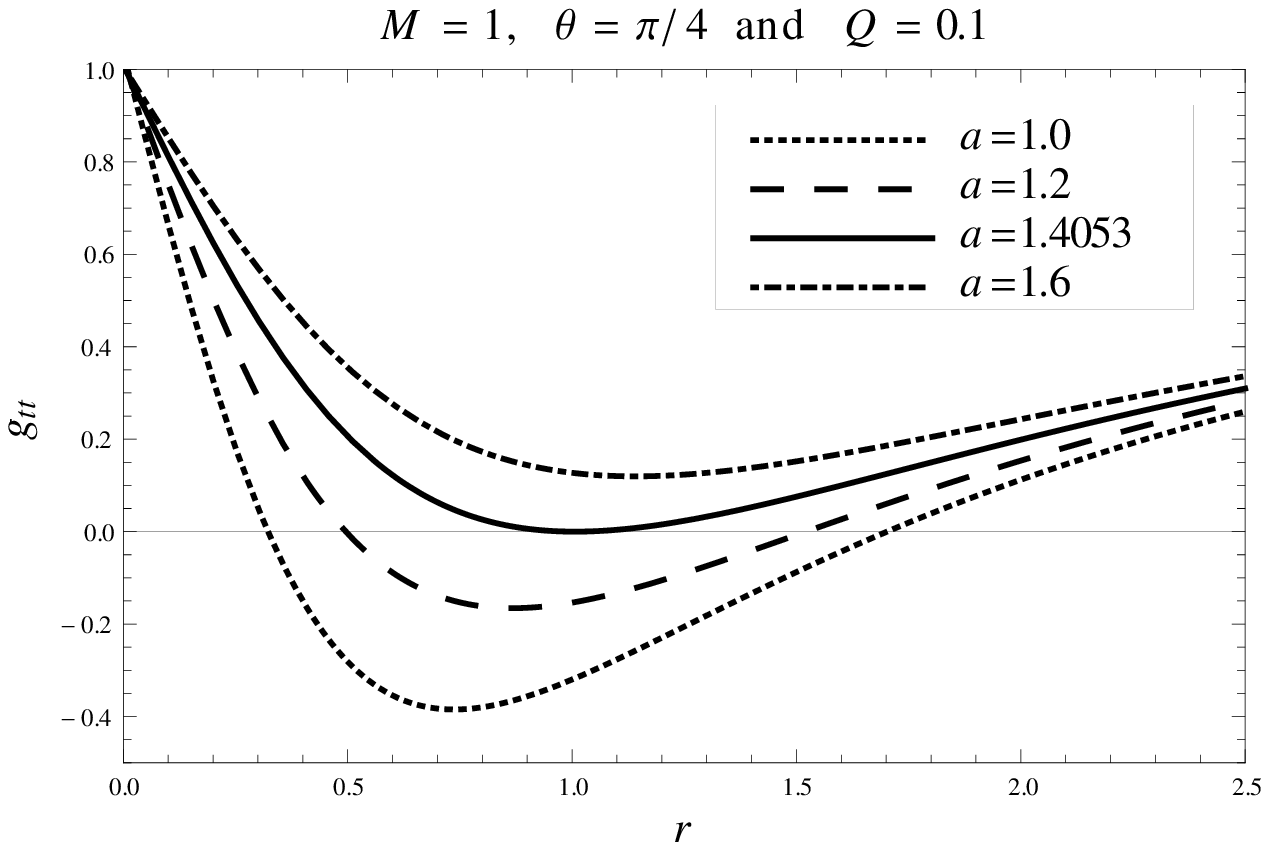}\hspace{-1cm}
 &\includegraphics[scale=0.65]{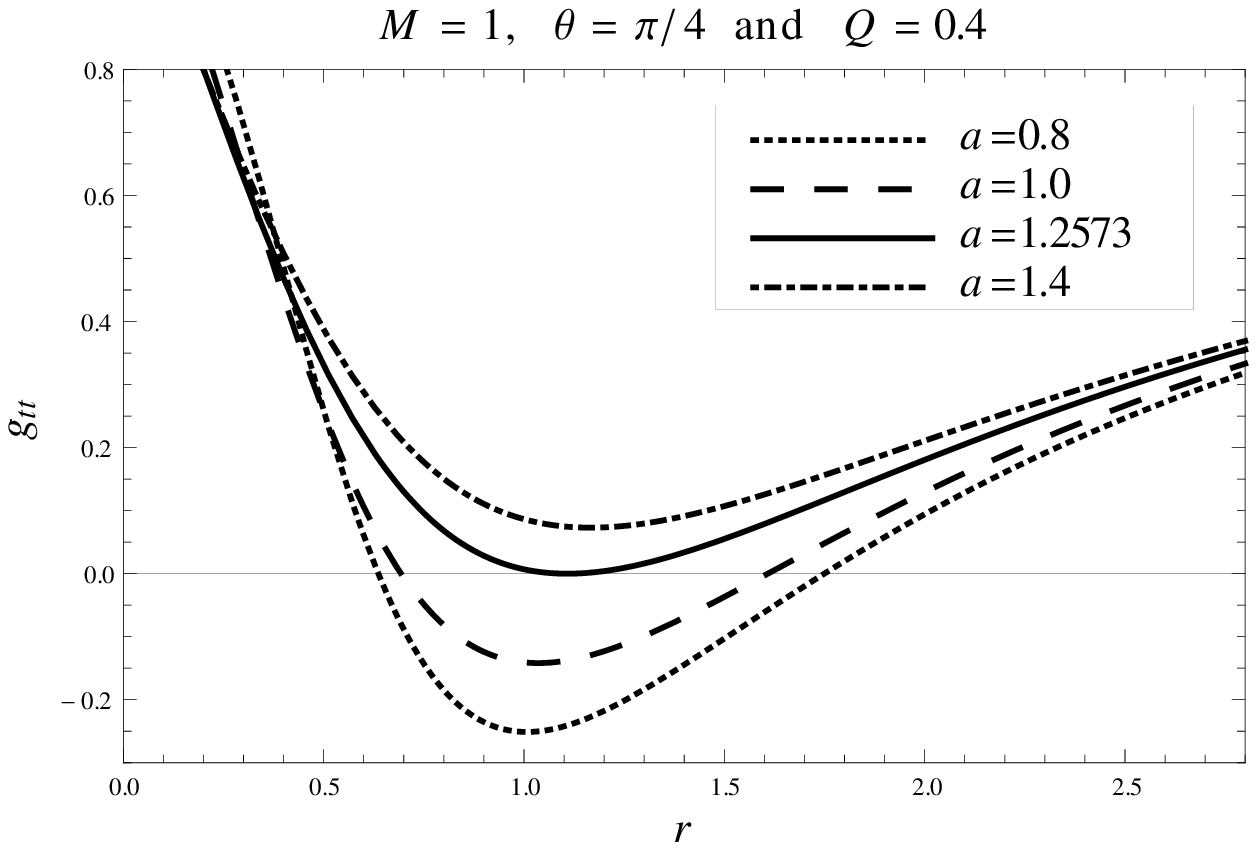}
 \end{tabular}
 \caption{The behavior of static limit surface ($g_{tt}$) vs $r$ for 
 different values of $a$. The possibility of existence of horizon is 
 decreases with increases values of $Q$ and $a$.}\label{fig1}
\end{figure*}
\begin{figure*}
\begin{tabular}{c c}
\includegraphics[scale=0.65]{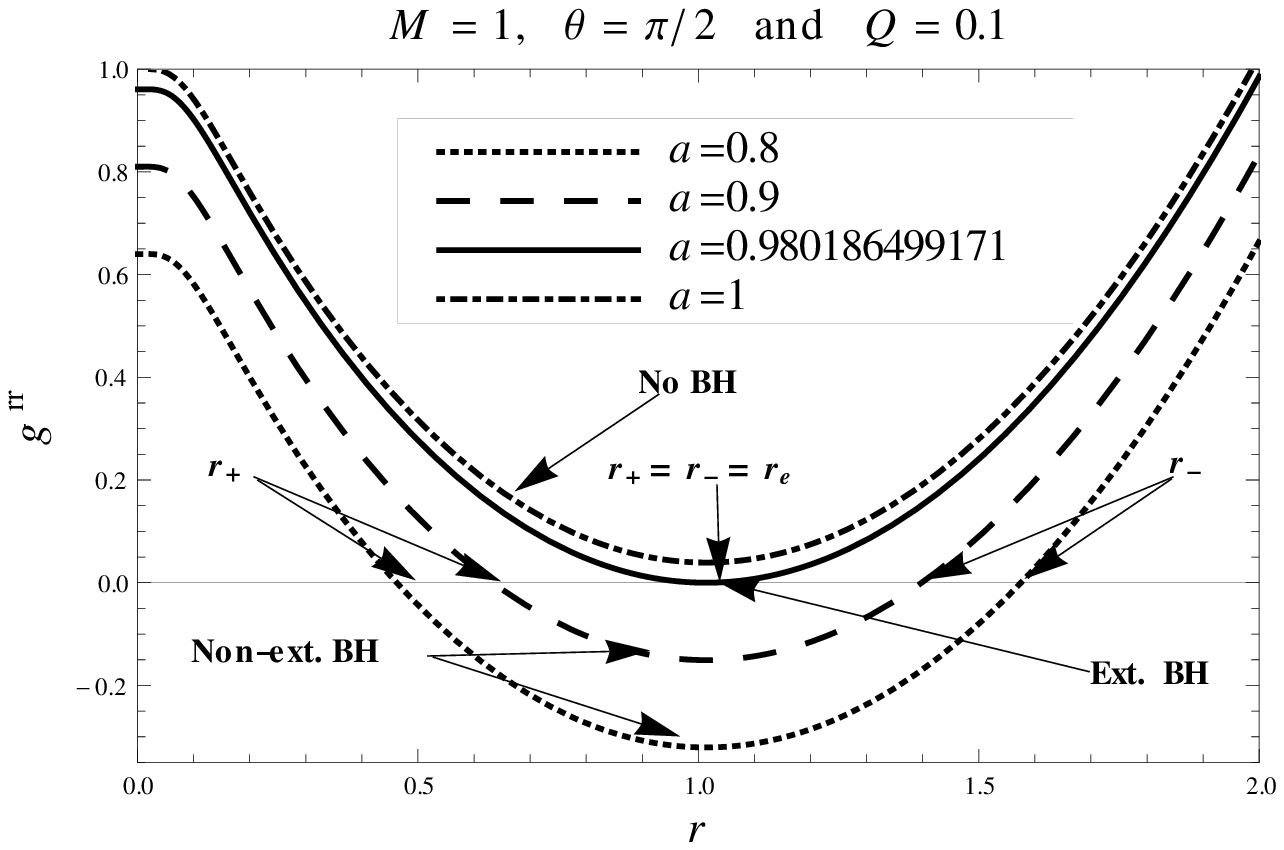}\hspace{-1cm}
&\includegraphics[scale=0.65]{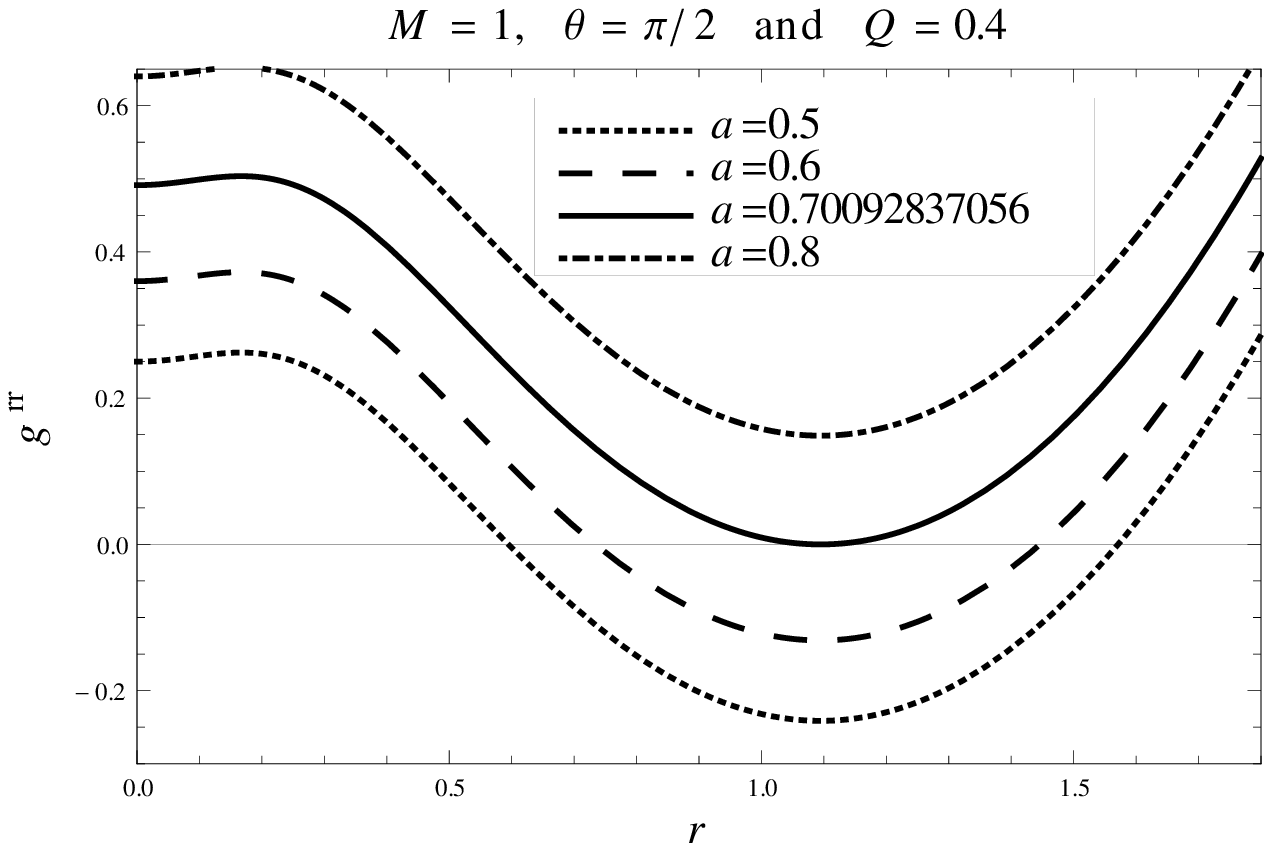}
\end{tabular}
 \caption{The behavior of $g^{rr}$ vs $r$ for different values of $a$. Panel (a) for $Q=0.1$. Panel (b) for $Q=0.4$.}\label{fig2}
\end{figure*}
\begin{figure*}
\begin{tabular}{c c}
\includegraphics[scale=0.65]{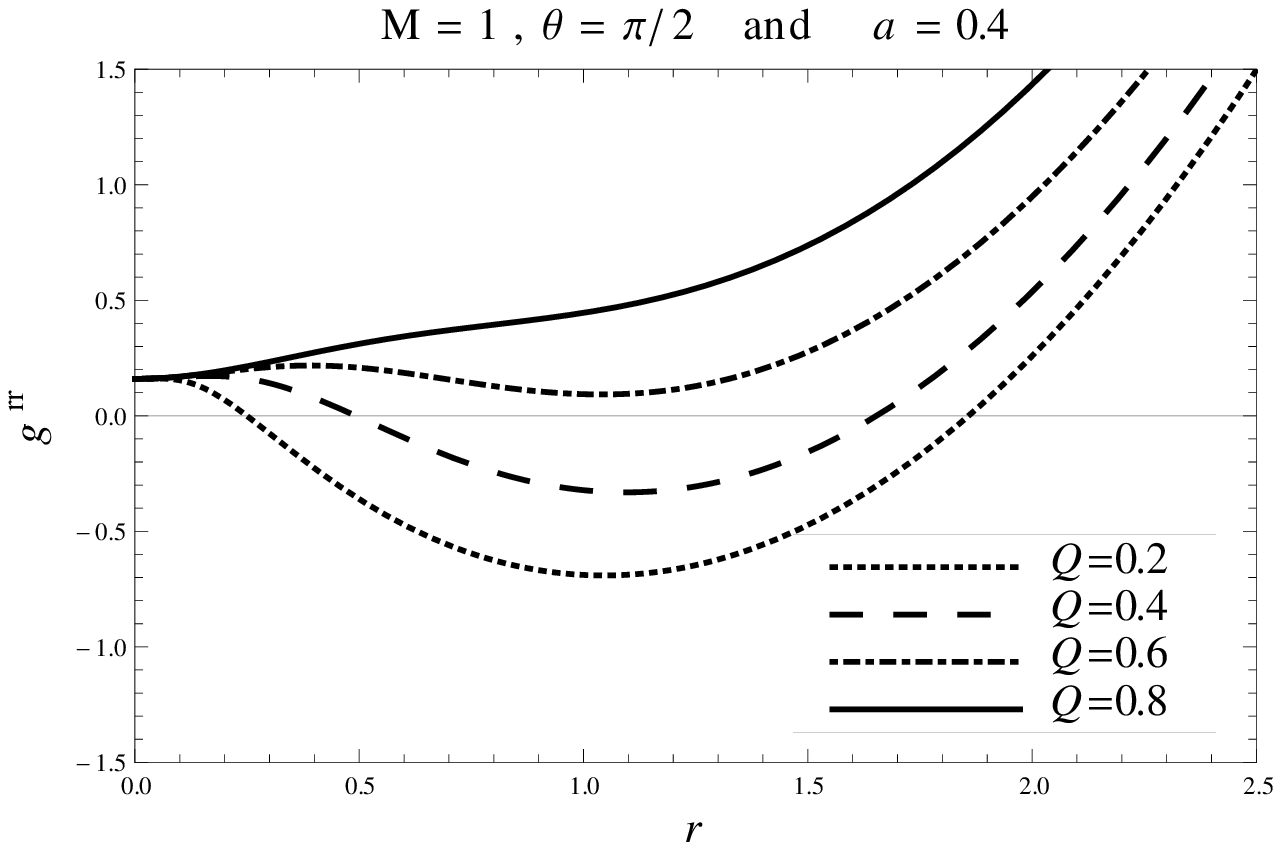}\hspace{-1cm}
&\includegraphics[scale=0.65]{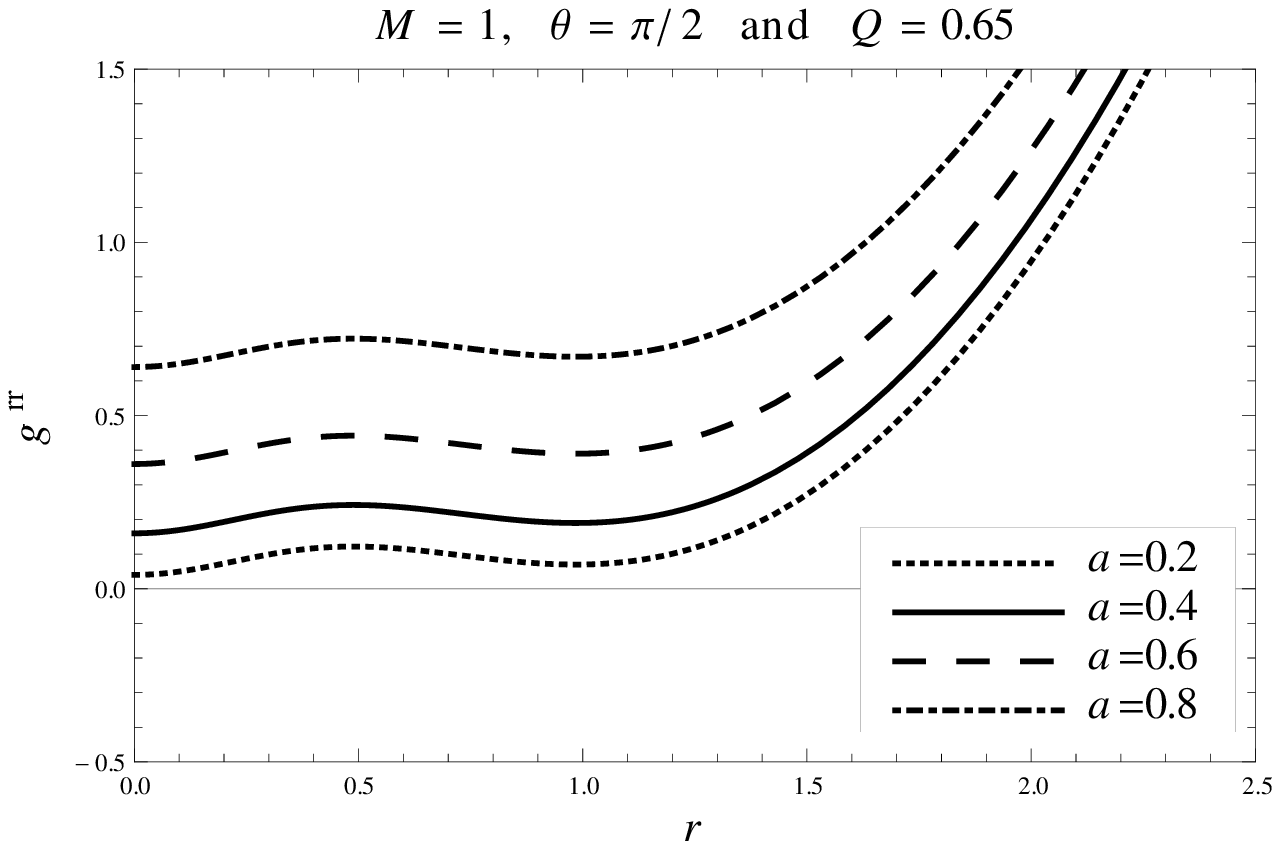}
\end{tabular}
 \caption{The behavior of $g^{rr}$ vs $r$. Panel (a) for fixed $a=0.4$ and different 
 values of $Q$. Panel (b) for fixed $Q=0.65$ and different values of $a$.}\label{fig3}
\end{figure*}
\begin{figure*}
\begin{tabular}{c c}
\includegraphics[scale=0.6]{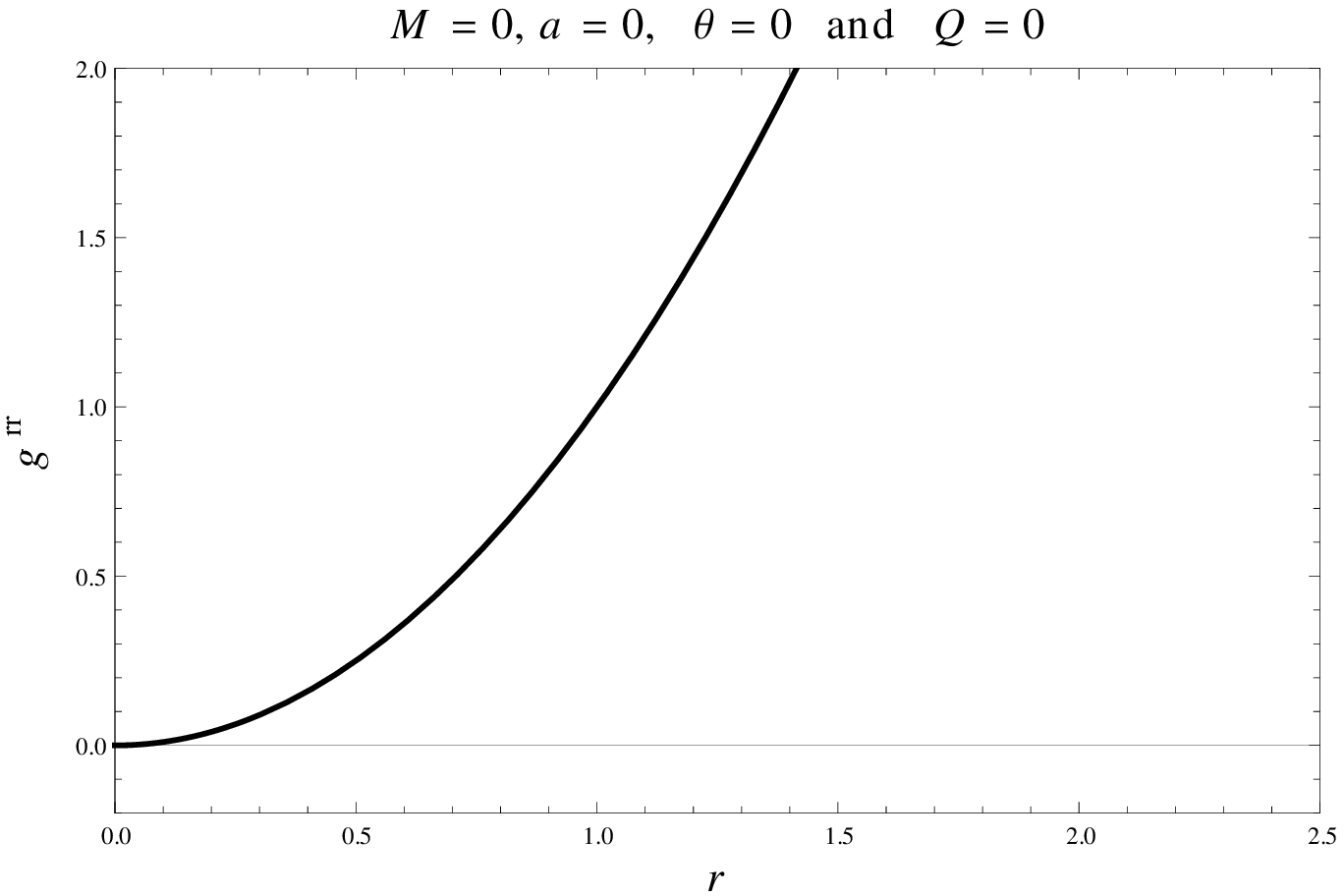}\hspace{-0.5cm}
&\includegraphics[scale=0.63]{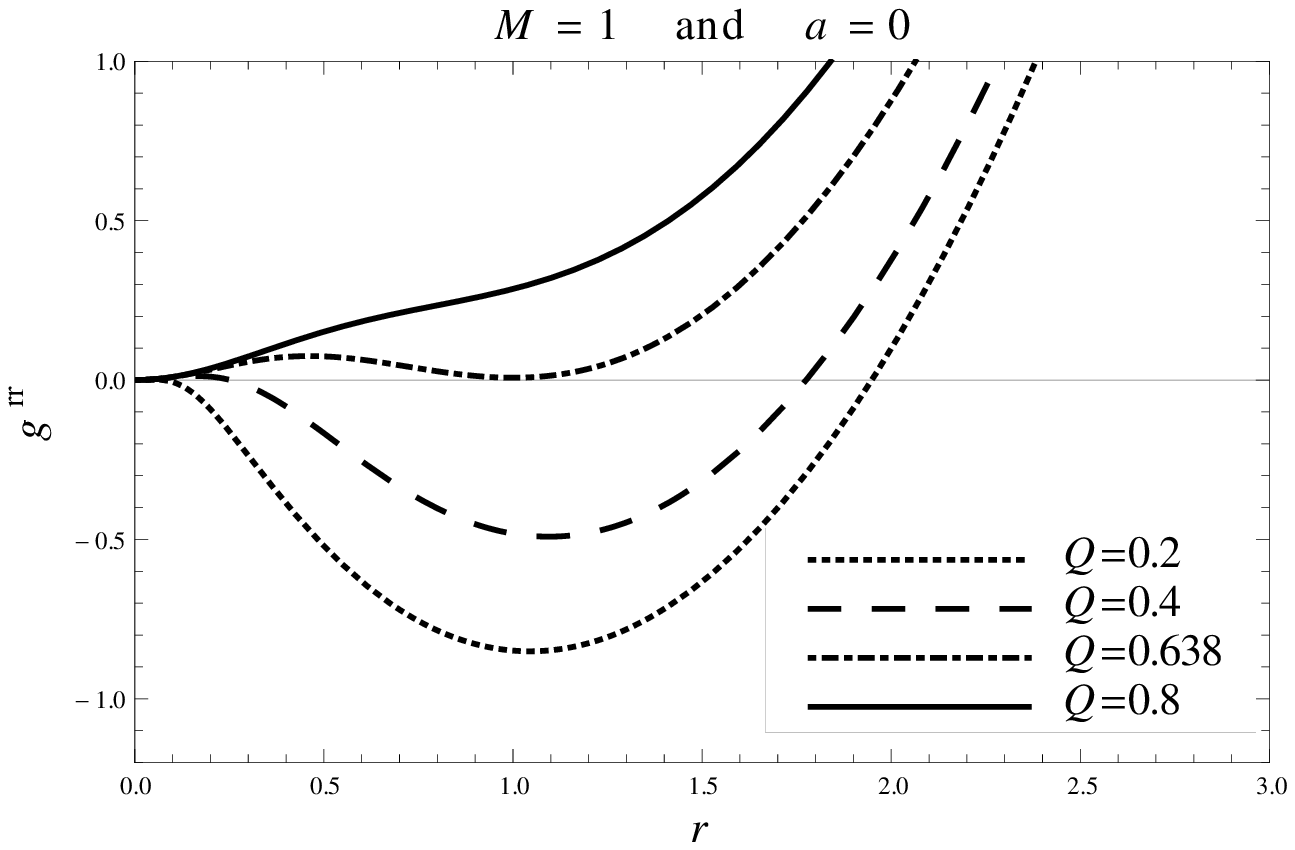}
\end{tabular}
 \caption{The behavior of $g^{rr}$ vs $r$. Panel (a) for $a=M=Q=0$. Panel (b) 
 for $M=1$, $a=0$ and different values of $Q$.}\label{fig4}
\end{figure*}
\begin{equation}\label{eh}
\Delta = \Sigma f(r,\theta)+a^2 \sin^2 \theta =0,
\end{equation}
which leads to
\begin{eqnarray}\label{eh1}
\Sigma(\Sigma+Q^2)^2-2Mr\Sigma^{3/2}(\Sigma+Q^2)^{1/2}+Q^2\Sigma^2 
\nonumber && \\+a^2(\Sigma+Q^2)^2 \sin^2 \theta = 0.
\end{eqnarray}
Clearly, the radii of the horizons depends on $\theta$, which are different from that in the usual Kerr case.
The behavior of the static limit surface is shown in Fig.~\ref{fig1} and that of event horizon in 
Fig.~\ref{fig2}-\ref{fig4}  for different values of mass $M$, charge $Q$ and spinning parameter $a$.
The two figures reveals that there exists set of values of parameters for which we have two horizons, 
i.e., a regular BH with both inner and outer horizons.  One can also find values of parameters for which
one get extreme BHs where the two horizons are coincides. The region between the static limit surface and 
the event horizon is termed as quantum ergosphere. The ergosphere is the region which lies outside the BH.
In ergosphere it is possible to enter and leave again, and the object moves in the direction of the spin of BH.   
\section{Particle orbits in the  rotating Ay\'{o}n-Beato-Garc\'{i}a black hole }
\begin{figure}[h!]
\hspace{-0.4cm}
\includegraphics[scale=0.55]{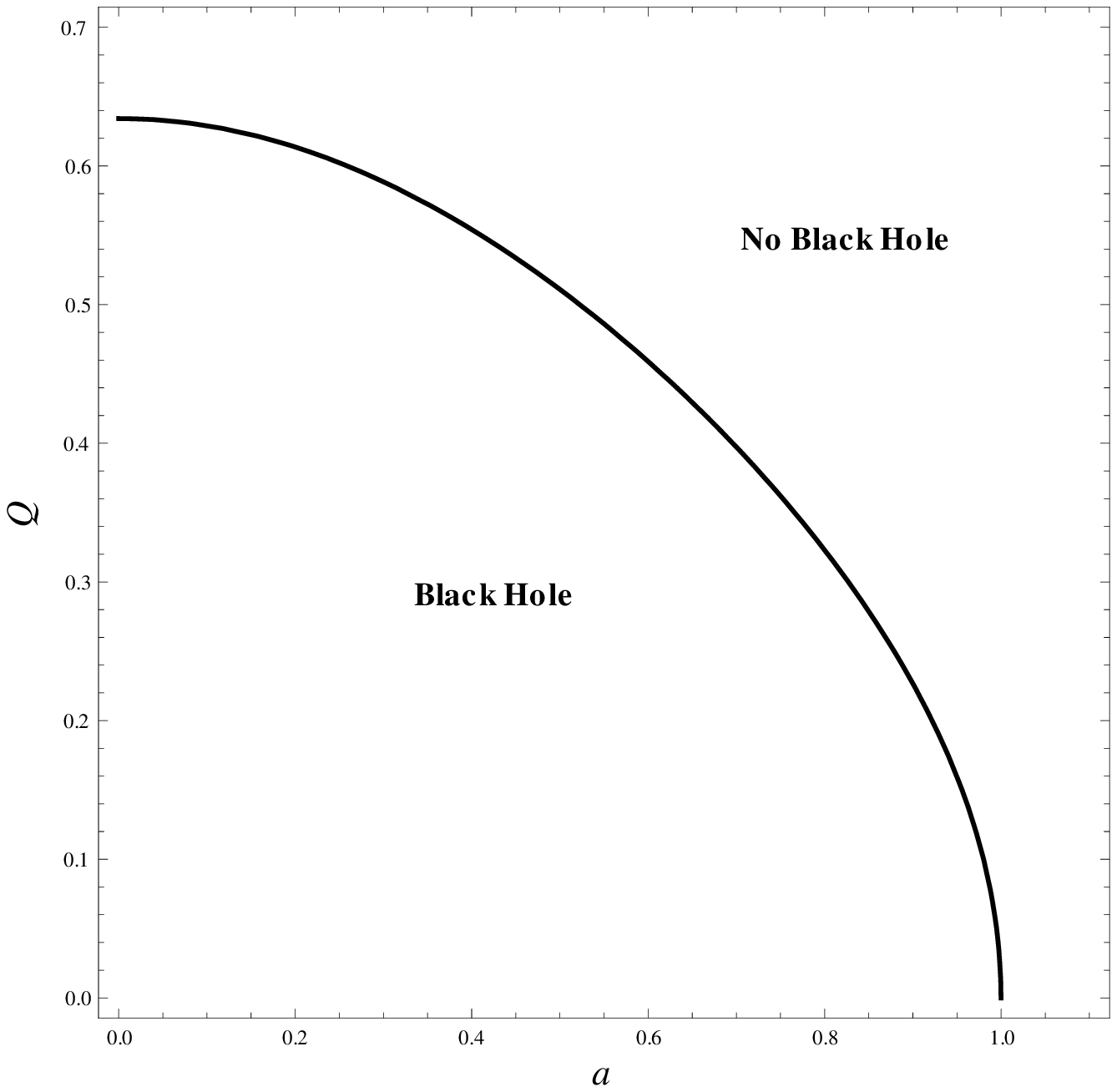}\hspace{-0.6cm}
\caption{Plot showing spin deformation parameter plane of rotating ABG metric.
The curve separates BH from no BH or configuration without an event horizon (i.e., no real root of $\Delta = 0$).}\label{figure5}
\end{figure}

Astrophysical BH candidates are supposed to be the Kerr BHs as predicted in general relativity, but the 
actual nature of these objects has still to be substantiated \cite{Bambi:2011mj,Bambi:2013qj,Zhang:1997dy}. 
If one wish to test the  nature of an astrophysical BH candidate, it is better to consider a more general spacetime, 
like rotating ABG BH, than a Kerr, in which the central object is described by a mass $ M $, spin parameter $ a $, and 
an additional deviation parameter $Q$.  The rotating ABG metric can be seen as the prototype of a large class of non-Kerr BH 
metrics and the  Kerr solution must be recovered when this deformation parameter $Q$ vanishes (cf. Fig.~\ref{figure5}). If 
the observations require vanishing deformation parameter, the compact object is a Kerr BH or non-Kerr BH  otherwise.  
In general,  observations allows both the possibility of a Kerr BH with a certain spin parameter and non-Kerr BH with 
different spin parameters.  Further, we are far from a reliable candidate for a quantum theory of gravity, hence recently 
more attention is given for  phenomenological approaches  to somehow solve these singularity problem in classical general 
relativity and study possible implications. In this context, an important line of research is represented by the work on 
the regular BH solutions  Hence, it pertinent to consider BSW mechanism for the rotating ABG black holes. 

Let us consider a motion for a particle with mass $m_{0}$ falling in the background of a rotating ABG BH.
The geodesic motion for this BH is determined by the following Hamilton-Jacobi equation
\begin{equation}\label{hje}
\frac{\partial S}{\partial \tau} = -\frac{1}{2} g^{\mu\nu} \frac{\partial S}{\partial x^{\mu}} \frac{\partial S}{\partial x^{\nu}},
\end{equation}
where $\tau$ is an affine parameter along the geodesics, and $S$ is the Jacobi action. For this BH 
background the Jacobi action $ S $ can be separated as
\begin{equation}\label{hja}
S = \frac{1}{2} m_0^2 \tau -Et + L \phi + S_{r}(r) + S_{\theta}(\theta),
\end{equation}
where $ S_{r} $ and $ S_{\theta} $ are function of $ r $ and $ \theta $ respectively. The constants 
$ m_0 $, $ E $, and $ L $ correspond to rest mass, conserved energy and angular momentum through
$m_{0}^2= -p_{\mu}p^{\mu}$, $E=-p_{t}$, and $L=p_{\phi}$. For equatorial plane ($\theta=\pi/2$) in 
the ABG metric (\ref{abg}), we obtain the null geodesics in the form of the first-order differential equations
\begin{equation}\label{u^t}
 \frac{d t}{d \tau} = \frac{1}{r^2} \Big[-a (aE - L) + \left(r^2 + a^2\right) \frac{T}{\Delta}\Big],
\end{equation}
\begin{equation}\label{u^Phi}
 \frac{d \phi}{d \tau} = \frac{1}{r^2} \Big[-(aE - L ) + \frac{a T}{\Delta}\Big], 
\end{equation}
\begin{equation}\label{u^r}
 \frac{d r}{d \tau} = \pm \frac{1}{r^2} \sqrt{T^2 -\Delta \left[m_0^2 r^2  + (L-a E)^2 + K\right]},
\end{equation}
where $ T = E (r^2 + a^2) -La $ and $K$ is a constant.  
\begin{figure*}
\begin{tabular}{c c}
\includegraphics[scale=0.7]{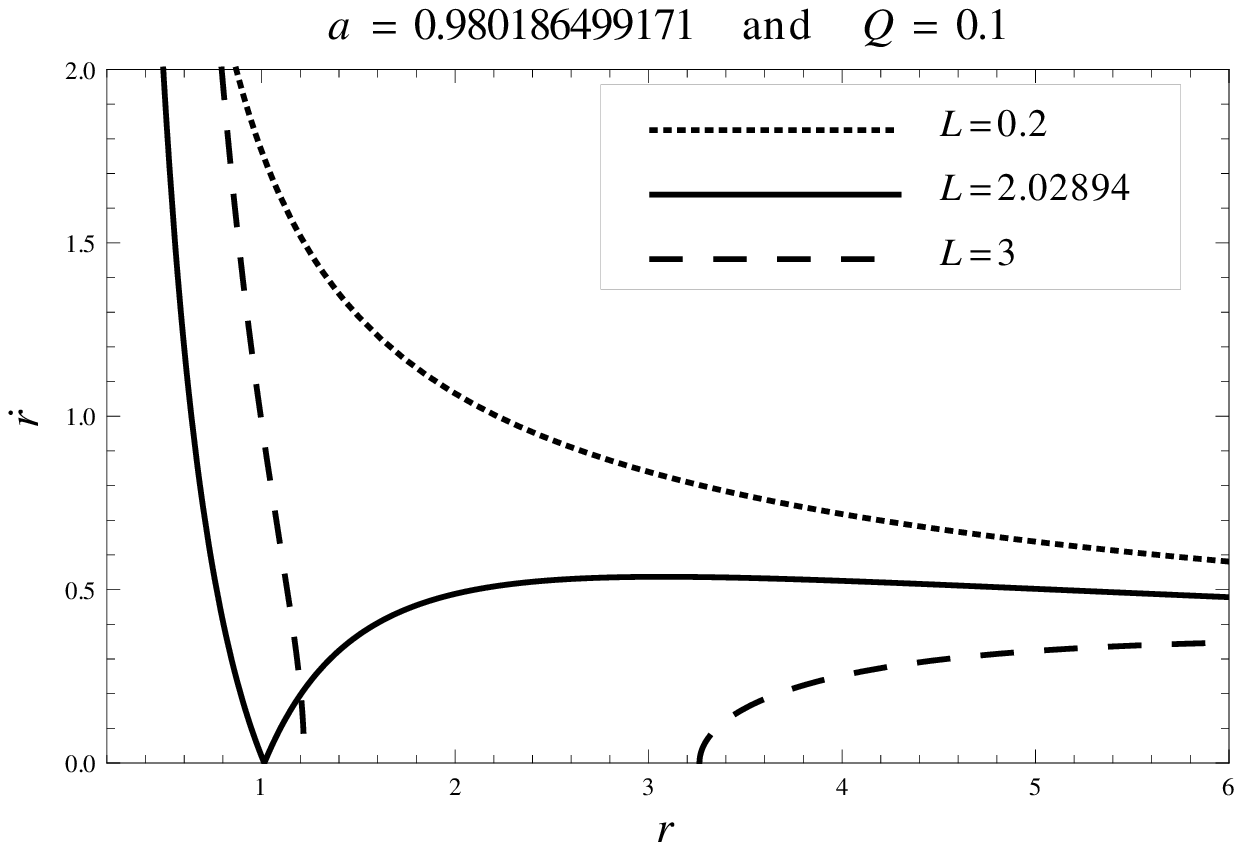}\hspace{-1cm}
&\includegraphics[scale=0.7]{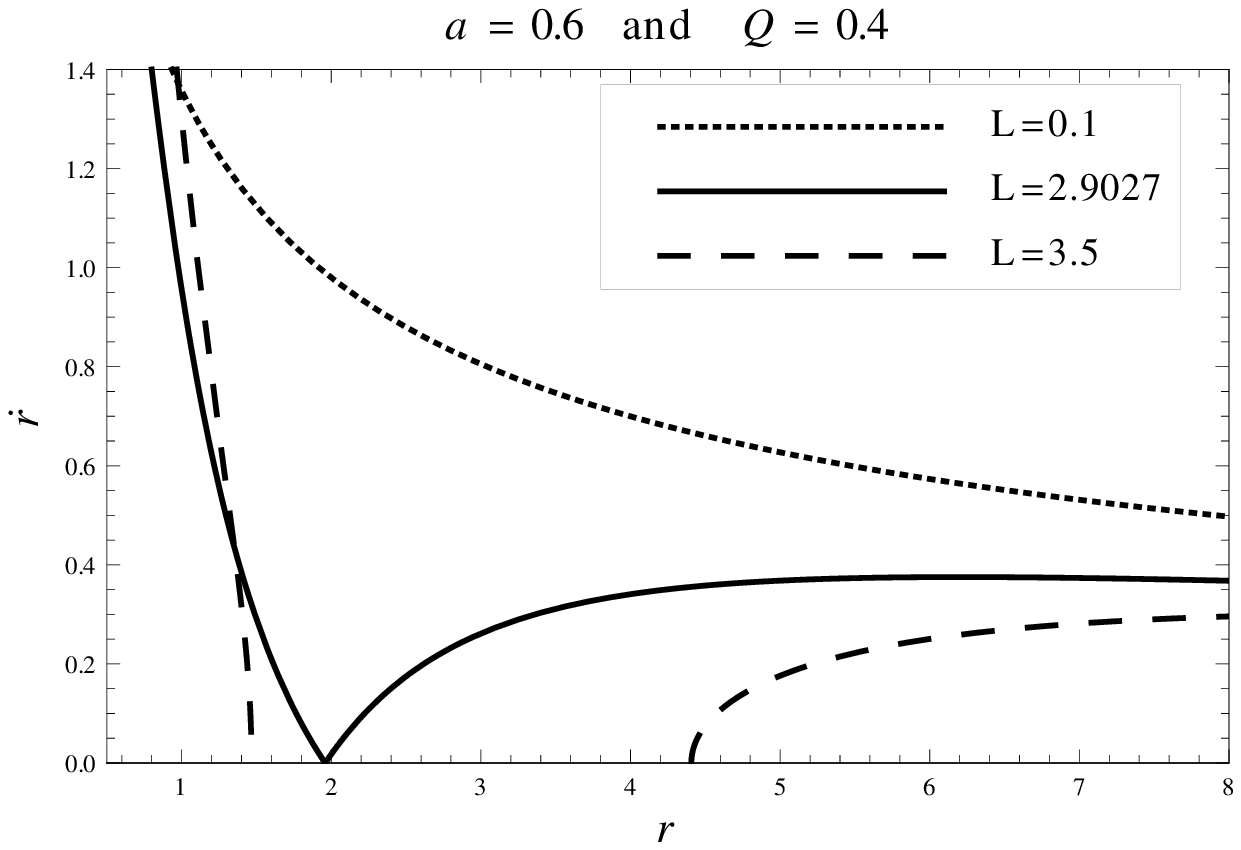}
\end{tabular}
 \caption{The behavior of $\dot{r}$ vs $r$ for different values of $L$. Panel (a) for extremal BH.
 Panel (b) for non-extremal BH.}\label{fig6}
\end{figure*}
These equations determine the propagation of light in the spacetime of the rotating ABG BH. Obviously, 
for $ Q = 0 $, it is just the null geodesic for the Kerr BH. The constant $K=0$ is the necessary and 
sufficient condition for particles motion initially in the equatorial plane to remain in the equatorial plane. 
Any particle which crosses the equatorial plane has $K>0$ \cite{bard}.

The study of effective potential is a very useful tool for describing the motion of test particles.
The radial equation for timelike particles moving along geodesic in the equatorial plane ($\theta=\pi/2$) is described by
\begin{equation}
\frac{1}{2} \dot{r}^2 + V_{eff} = 0,
\end{equation}
with the effective potential
\begin{equation}
 V_{eff} =  -\frac{[E (r^2 + a^2) -La]^2 -\Delta [m_0^2 r^2  + (L-a E)^2]}{2 r^4},
\end{equation}
The circular orbit for the particle is defined as
\begin{equation}\label{lim}
V_{eff}=0 \;\;\;\text{and}\;\;\; \frac{dV_{eff}}{dr}=0,
\end{equation}
Eq.~(\ref{lim}) leads to the limitation on the possible values of angular momentum of free falling 
particle and to achieve the horizon of the BH the angular momentum $L$ must be lying within the range
$-4.80898 \lesssim L\lesssim 2.02893 $ for extremal case. For extremal case 
$a \approx 0.980186499171$ and $r_{e} \approx 1.01388$ is event horizon. Whereas for the case of Kerr-Newman BH
the extremal value of spin parameters is $a \approx 0.9$, range of angular momentum is
$-4.6864 \lesssim L \lesssim 2.0111$ and the $r_{e} \approx 1$. 

Since, $u^{t}= dt/d\tau \geq 0$, then from Eq.~(\ref{u^t}) the condition
\begin{eqnarray}
&& E \Big[(r^2+a^2)(r^2+Q^2)^2+2Mr^2a^2\sqrt{r^2+Q^2}-Q^2r^2a^2\Big] \nonumber \\
&& \;\;\;\;\;\;\;\;\;\;\;\;\;\geq Lar^2\Big[2M\sqrt{r^2+Q^2}-Q^2\Big],
\end{eqnarray}
must be satisfied, as $r \rightarrow r_{+}$, this condition reduces to
\begin{equation}
E \geq \frac{aL}{r_{+}^2+a^2} = \Omega_{+}L.
\end{equation}
\begin{figure*}
\begin{tabular}{c c}
\includegraphics[scale=0.7]{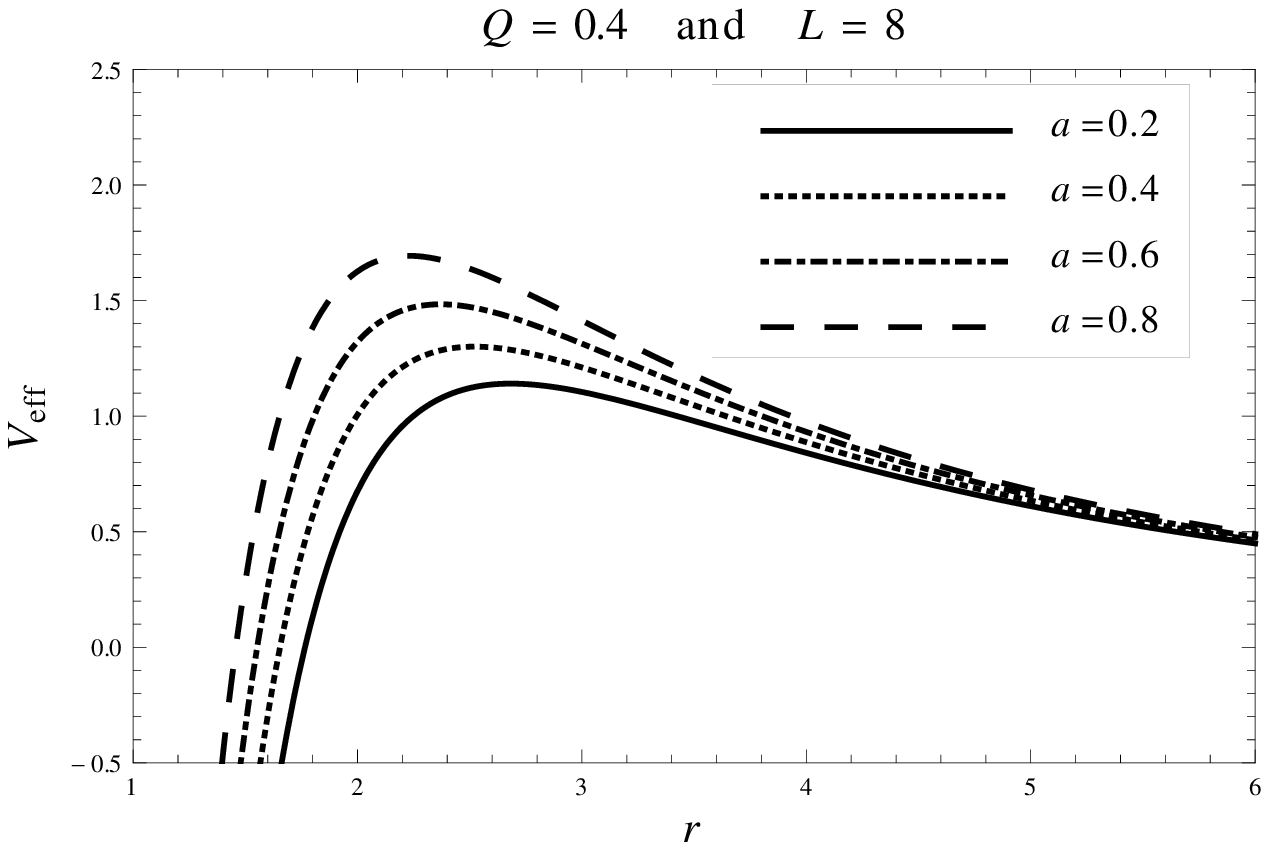}\hspace{-1cm}
&\includegraphics[scale=0.7]{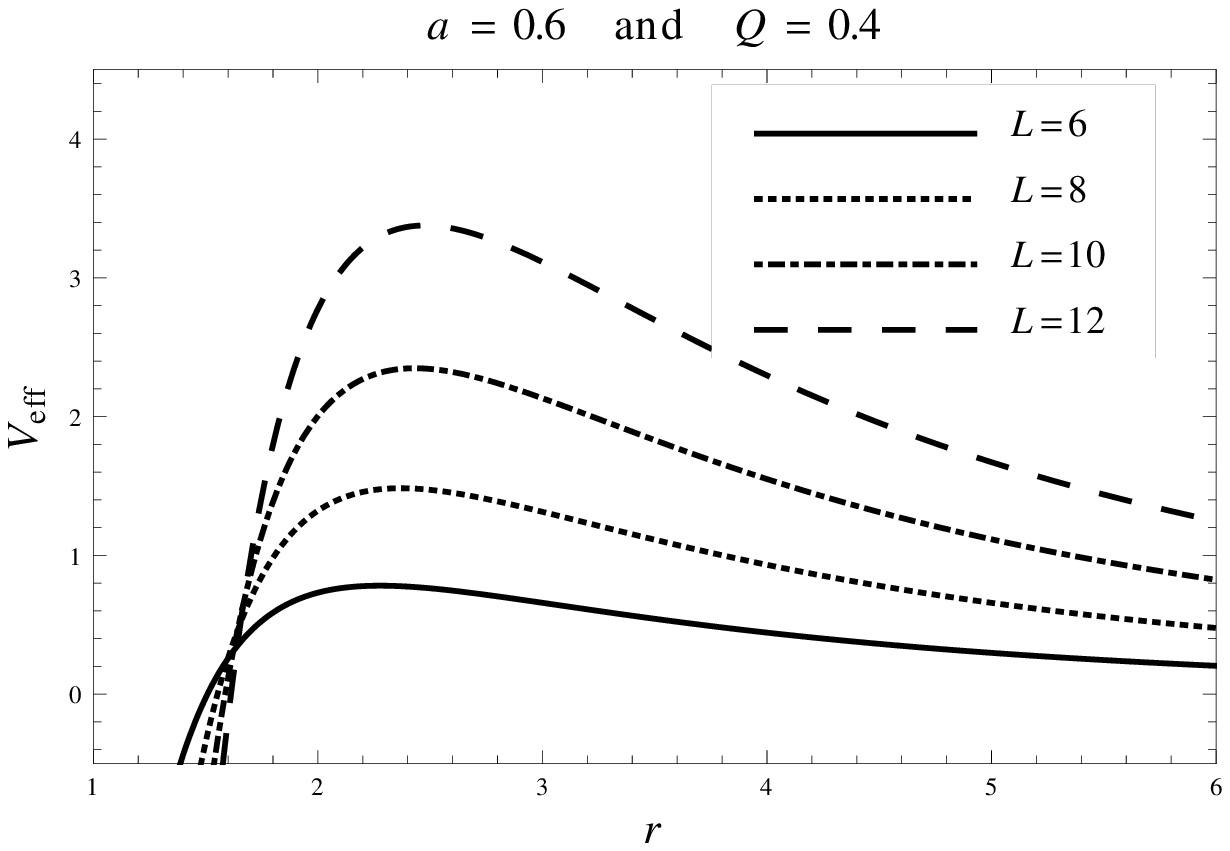}
\end{tabular}
 \caption{The behavior of $V_{eff}$ vs $r$. Panel (a) for $Q=0.4$, $L=8$ and different 
 values of $a$. Panel (b) for $Q=0.4$, $a=0.6$ and different values of $L$.}\label{fig7}
\end{figure*}
\section{Center of mass energy of two colliding particles}
We have dealt with range of the angular momentum, 
for which the particle can reach the horizon, i.e., if the angular momentum of 
particles are in the desired range, the collision can take place near horizon of the BH. 
We are now in the position to study the properties of the CM energy of two colliding 
particles with same rest mass $m_{0}$ near horizon of regular ABG BH.

\begin{figure*}
\begin{tabular}{c c}
\includegraphics[scale=0.7]{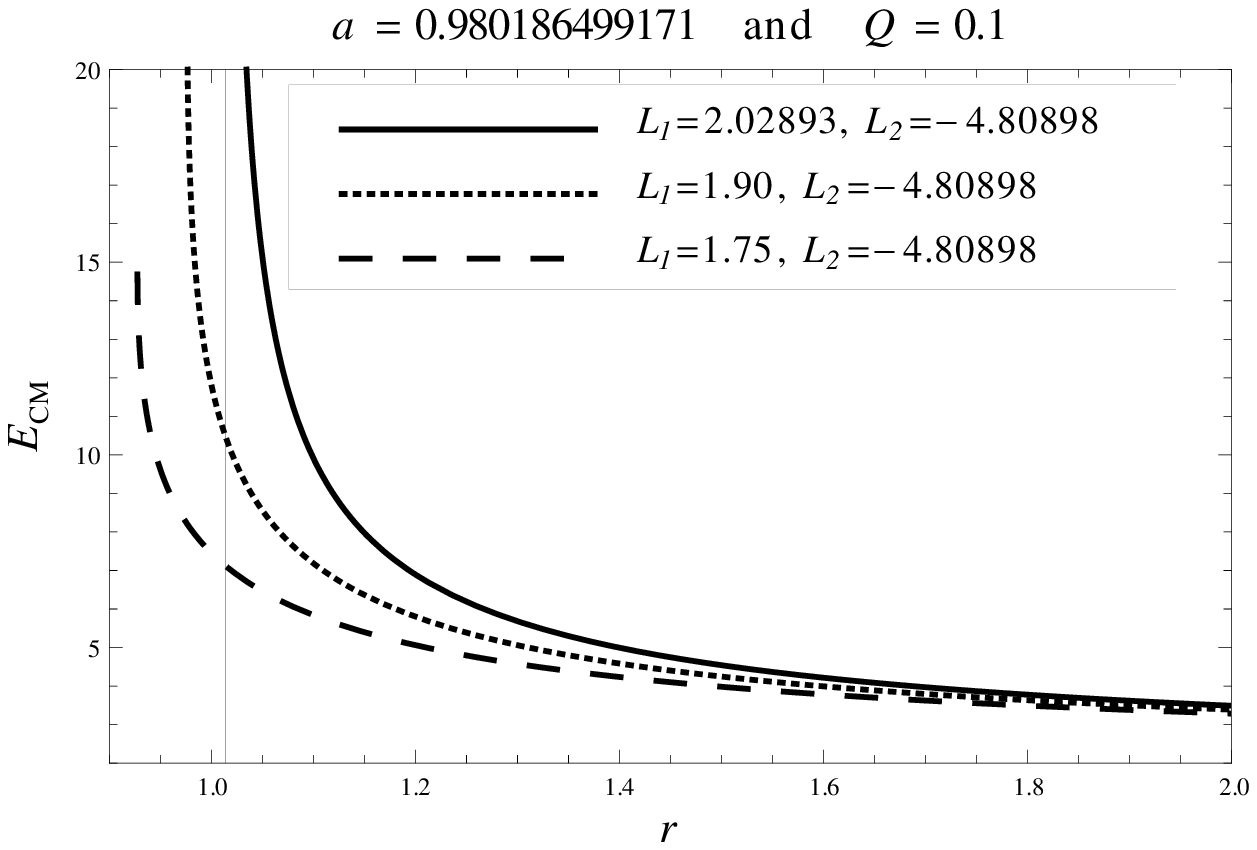}\hspace{-1cm}
&\includegraphics[scale=0.7]{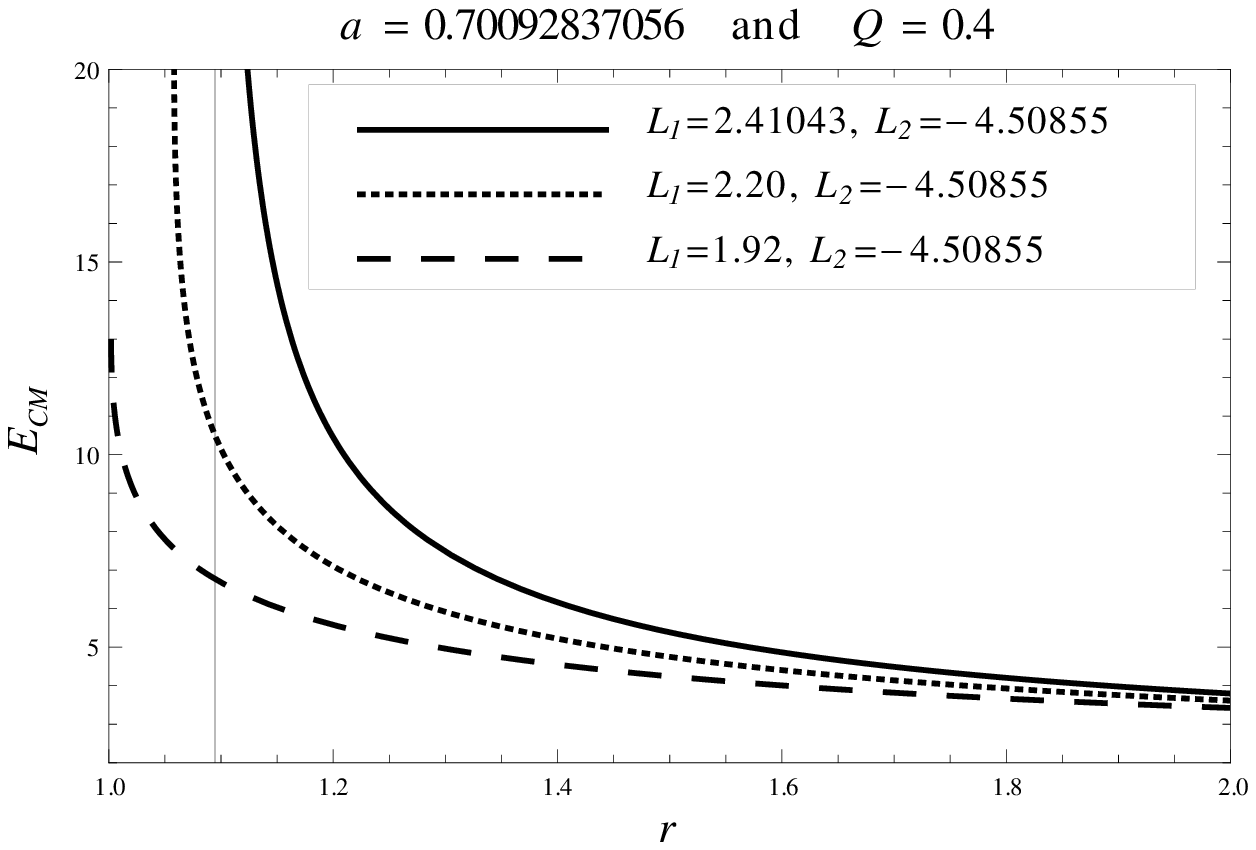}
\end{tabular}
 \caption{The behavior of $E_{CM}$ vs $r$ for extremal BH. Panel (a) for $a=0.980186499171$ and $Q=0.1$. 
 Panel (b) for extremal values of $a=0.70092837056$ and $Q=0.4$ where the vertical lines denote the location
 of event horizon.}\label{fig8}
\end{figure*}
\begin{figure*}
\begin{tabular}{c c}
\includegraphics[scale=0.7]{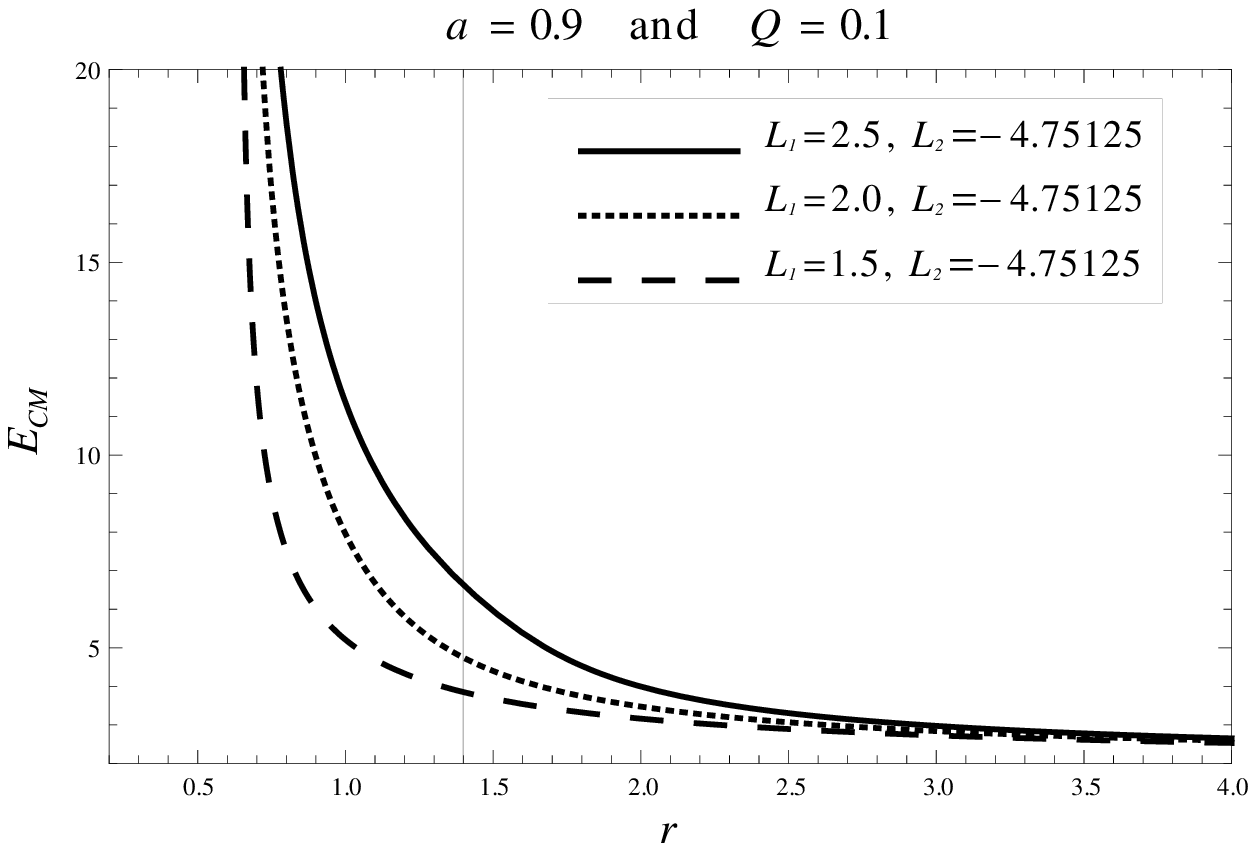}\hspace{-1cm}
&\includegraphics[scale=0.7]{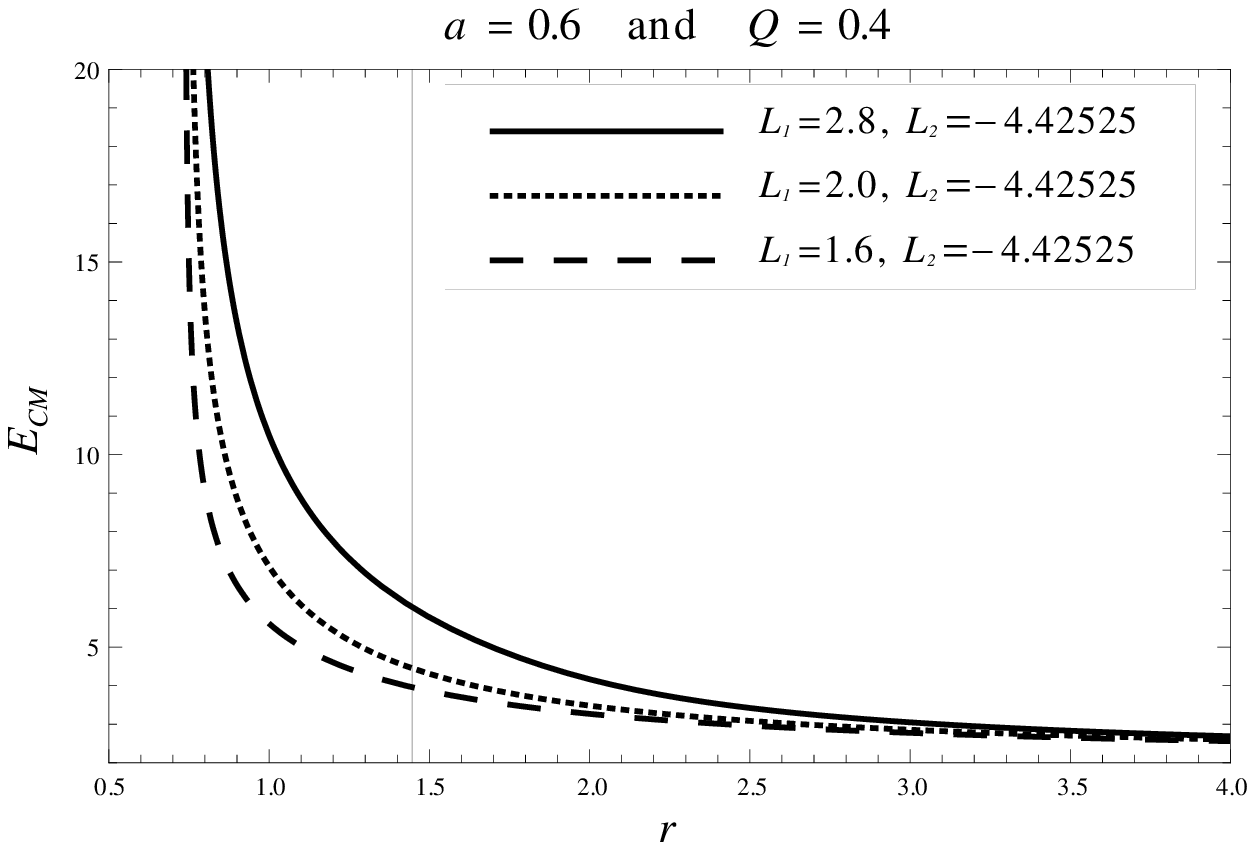}
\end{tabular}
 \caption{The behavior of $E_{CM}$ vs $r$ for non-extremal BH. Panel (a) for $a=0.9$ and $Q=0.1$.
 Panel (b) for $a=0.6$ and $Q=0.4$ where the vertical lines denote the location of the event horizon.}\label{fig9}
\end{figure*}
\begin{figure*}
\begin{tabular}{c c}
\includegraphics[scale=0.7]{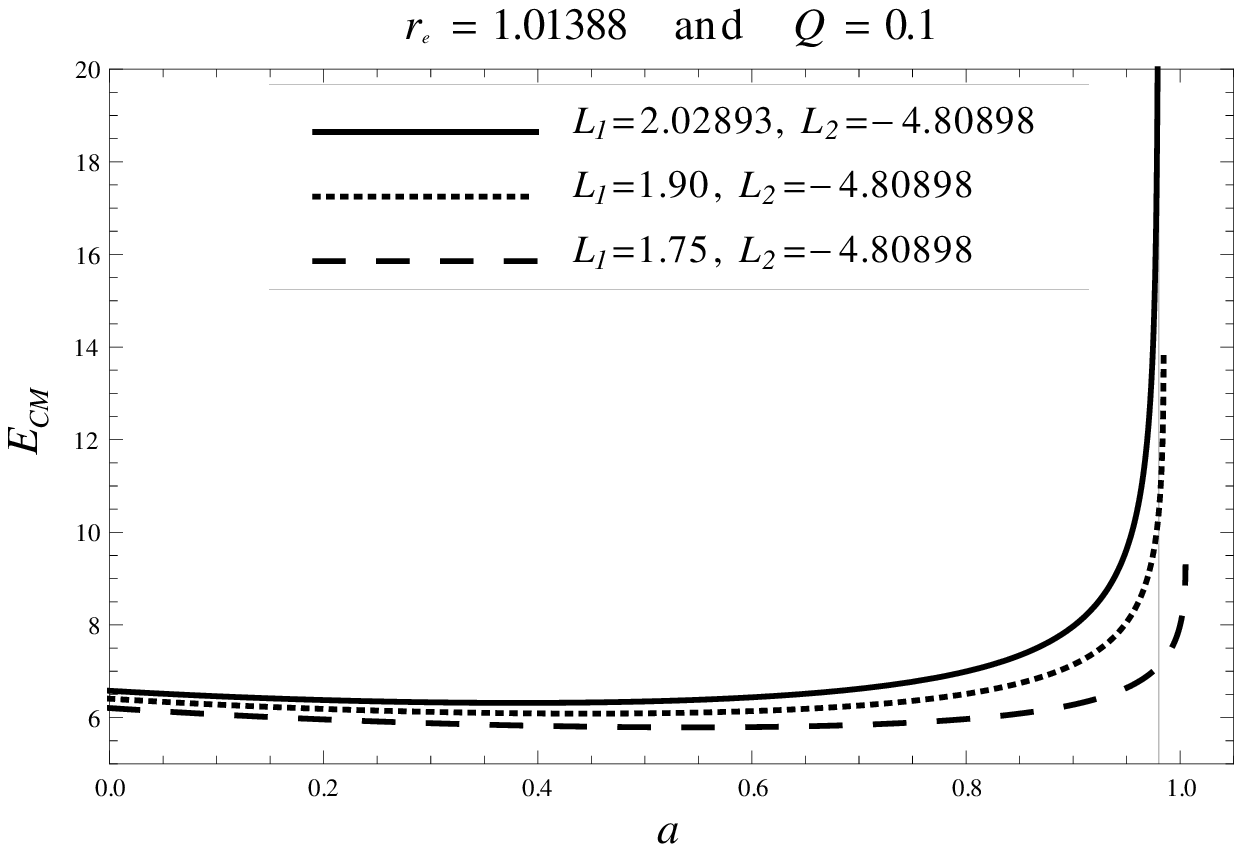}\hspace{-1cm}
&\includegraphics[scale=0.7]{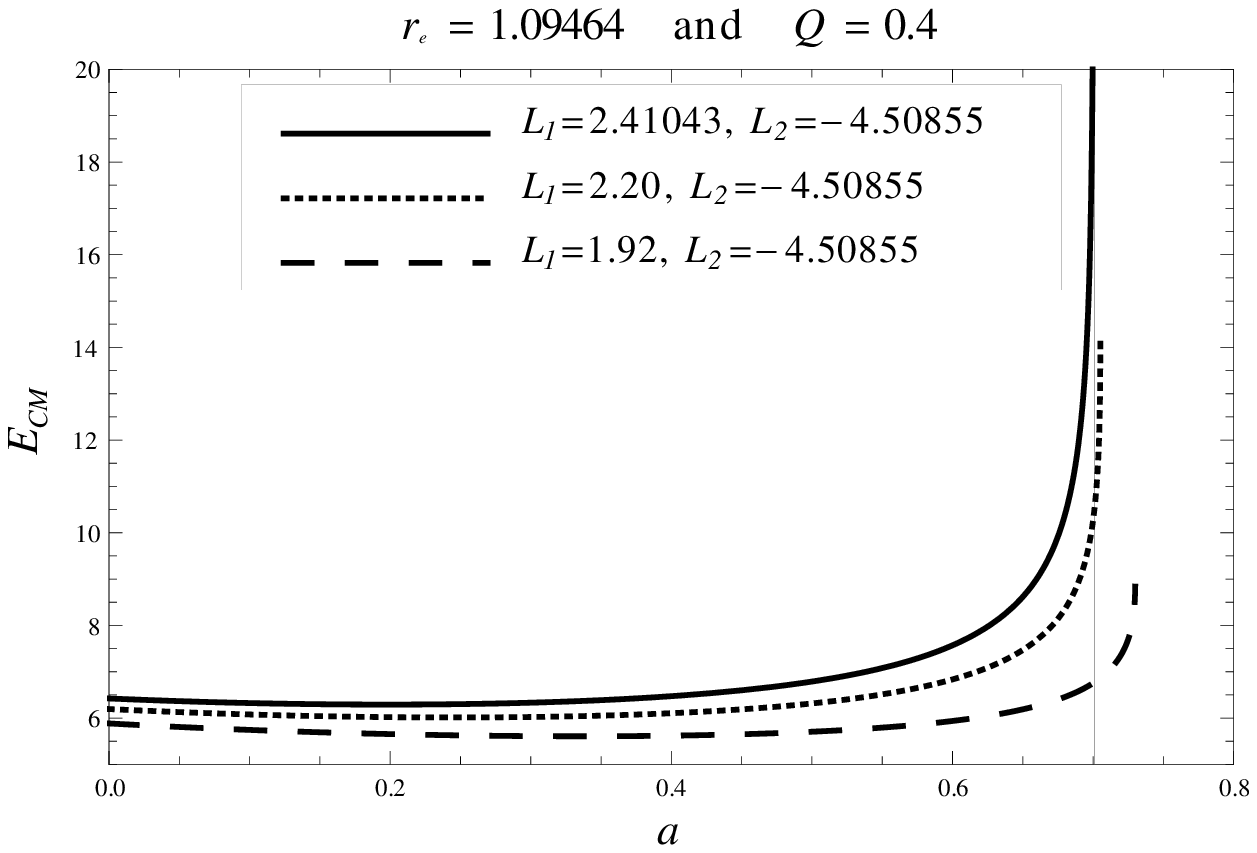}
\end{tabular}
 \caption{The behavior of $E_{CM}$ vs $a$ for extremal BH. Panel (a) for different values of $L_{1}$ where 
 vertical line denotes the location of  $a=0.980186499171$. Panel (b) for different values of $L_{1}$ where 
 vertical line denotes the location of  $a=0.70092837056$.}\label{fig10}
\end{figure*}
\begin{figure*}
\begin{tabular}{c c}
\includegraphics[scale=0.7]{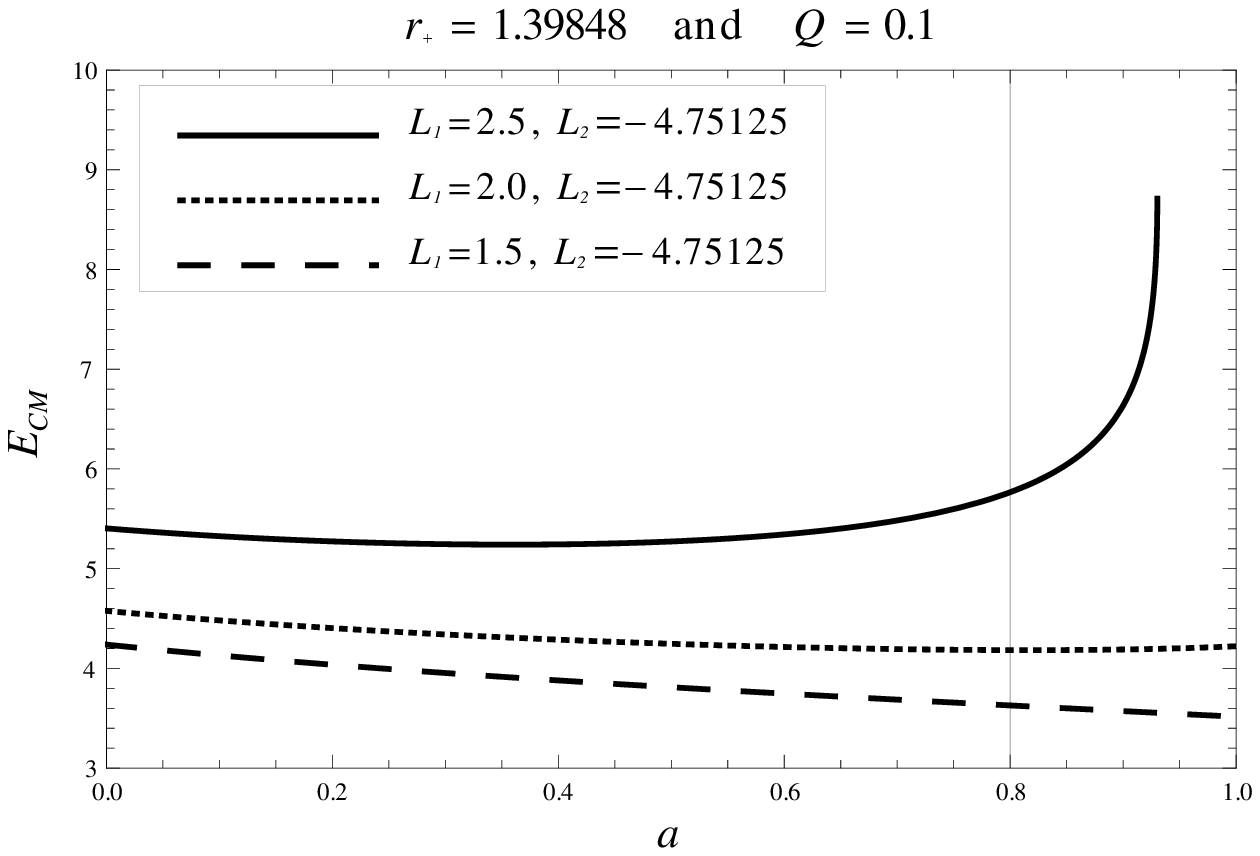}\hspace{-1cm}
&\includegraphics[scale=0.7]{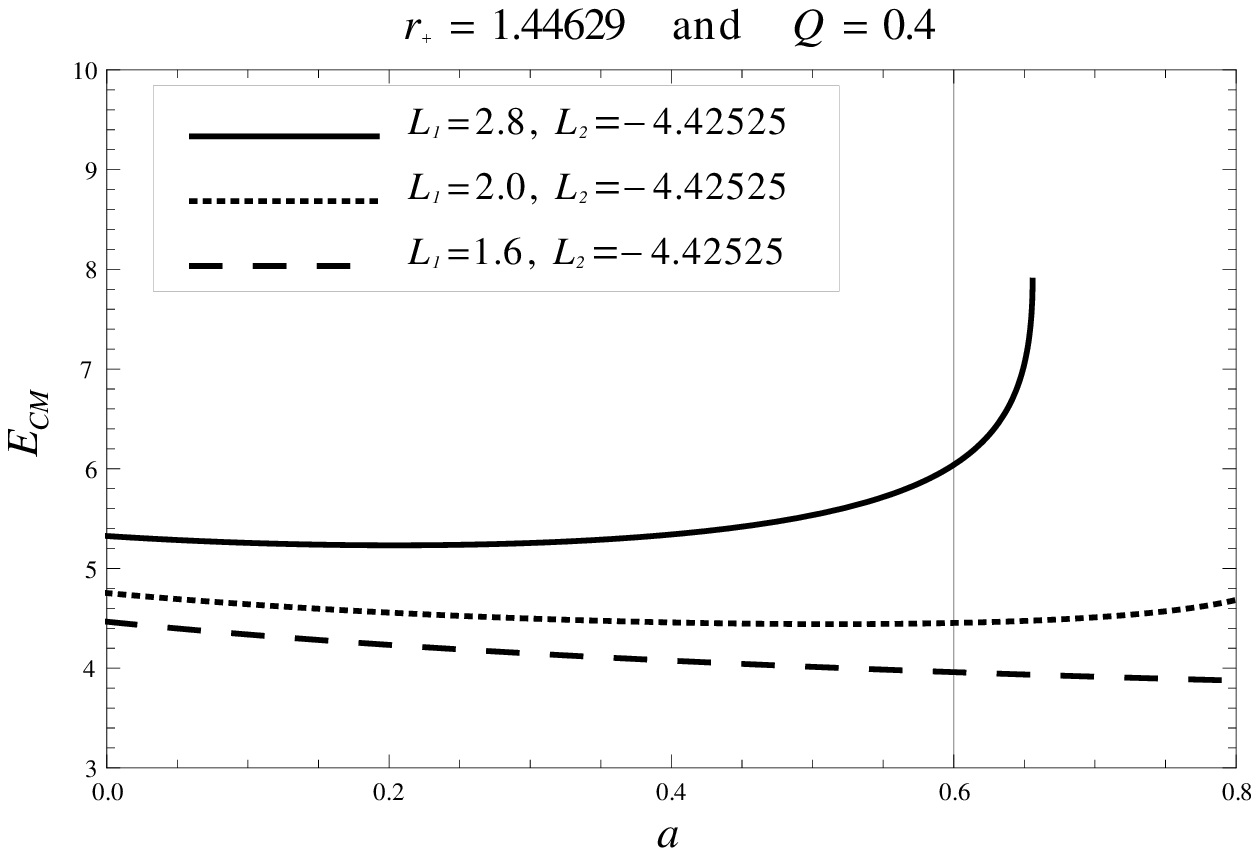}
\end{tabular}
 \caption{The behavior of $E_{CM}$ vs $a$ for non-extremal BH. Panel (a) for different values of $L_{1}$ where 
 vertical line denotes the location of  $a=0.9$. Panel (b) for different values of $L_{1}$ where 
 vertical line denotes the location of  $a=0.6$.}\label{fig11}
\end{figure*}

\begin{figure*}
\begin{tabular}{c c}
\includegraphics[scale=0.7]{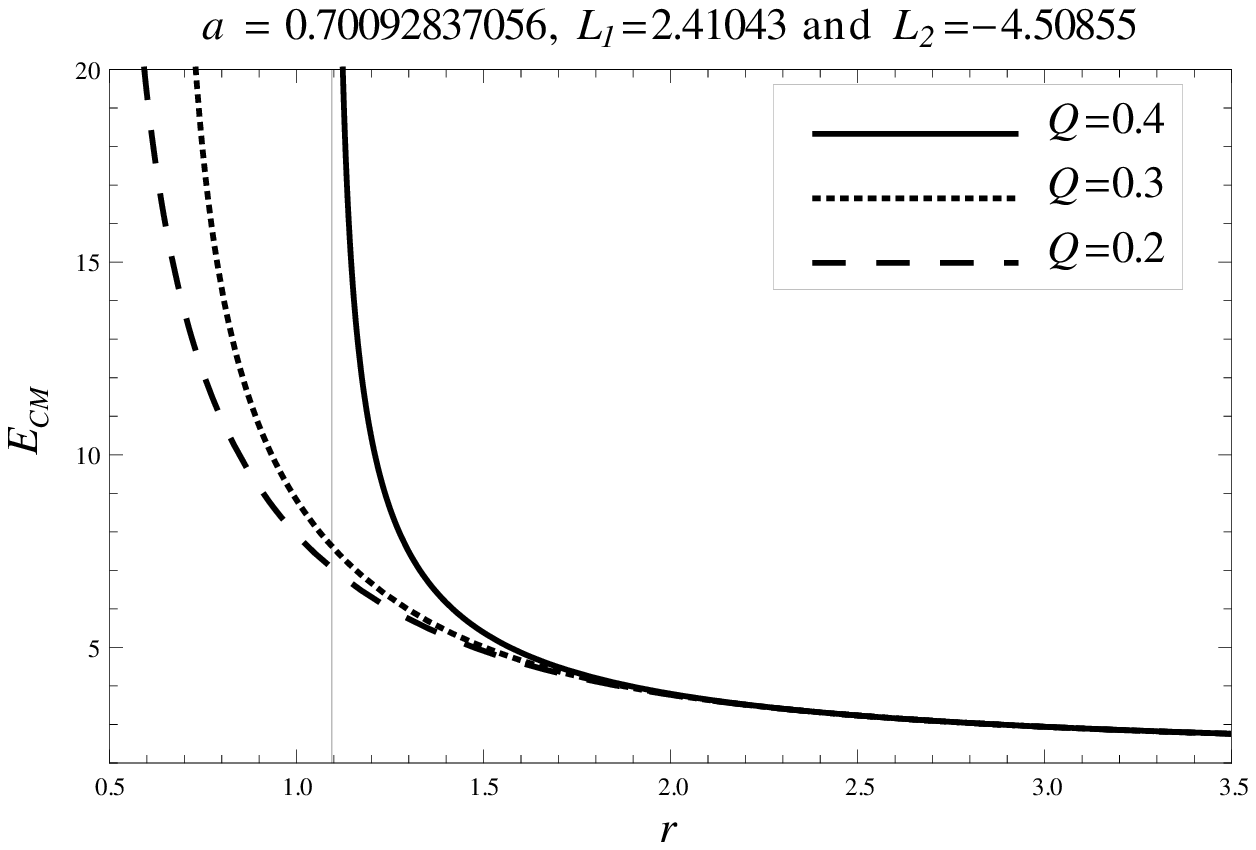}\hspace{-1cm}
&\includegraphics[scale=0.7]{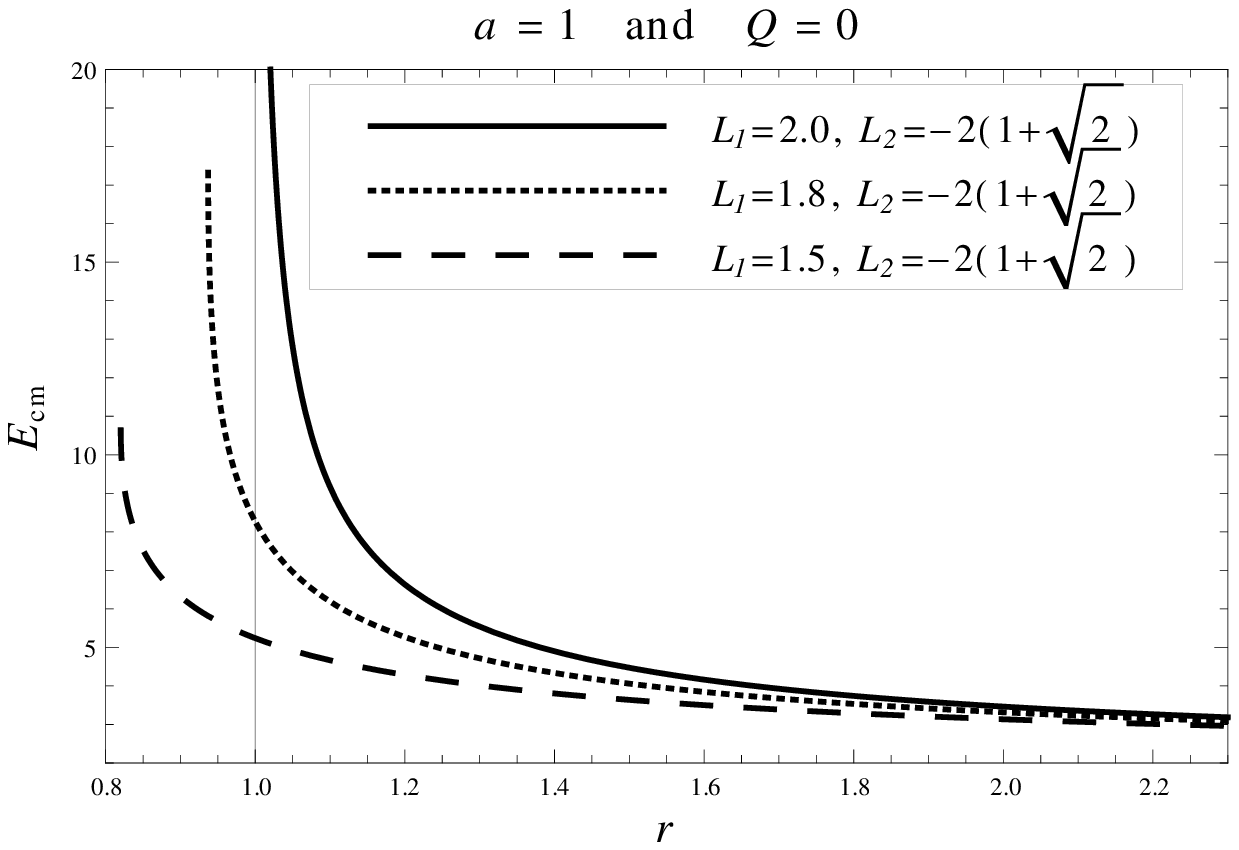}
\end{tabular}
 \caption{The behavior of $E_{CM}$ vs $r$. Panel (a) for different values of $Q$ where vertical line denotes the location of 
 $r=1.09464$. Panel (b) for extremal Kerr BH where vertical line denotes the location of event horizon.}\label{fig12}
\end{figure*}
\begin{figure*}
\begin{tabular}{c c}
\includegraphics[scale=0.7]{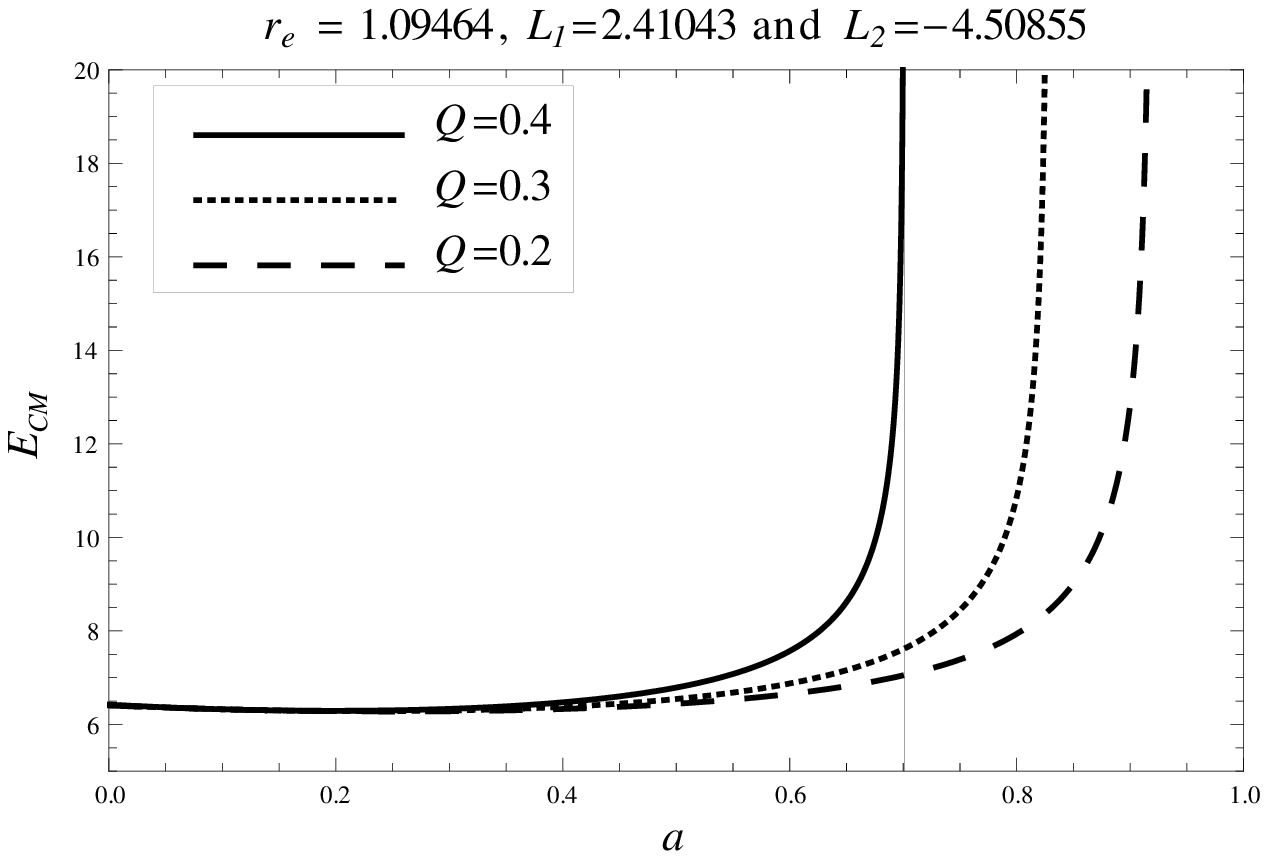}\hspace{-1cm}
&\includegraphics[scale=0.7]{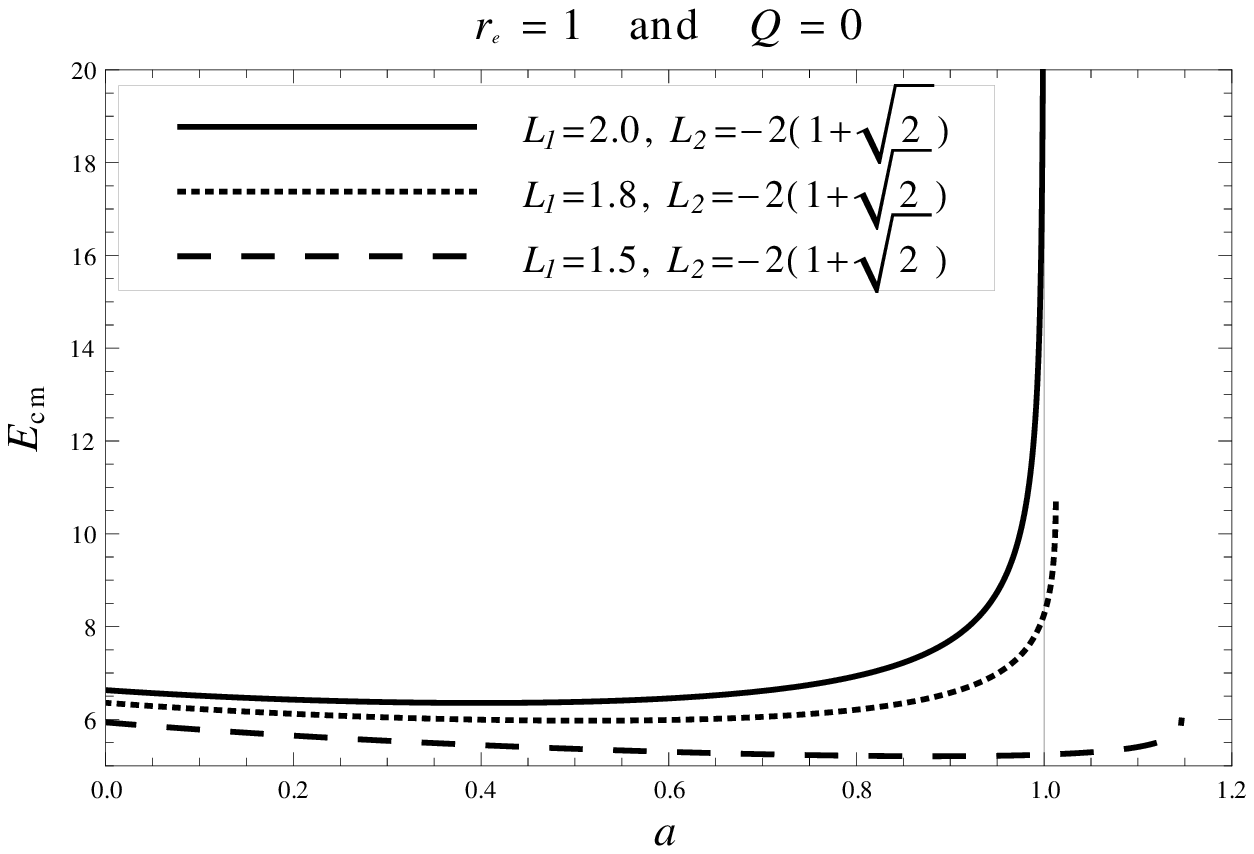}
\end{tabular}
 \caption{The behavior of $E_{CM}$ vs $a$. Panel (a) for different values of $Q$ where vertical line denotes the location of 
 $a=0.70092837056$. Panel (b) for extremal Kerr BH where vertical line denotes the location of 
 $a=1$.}\label{fig13}
\end{figure*}
Next, We study CM energy of the collision of two particles moving in equatorial plane of rotating ABG BH.
Let us consider two colliding particles with the same rest
mass $m_{1}= m_{2}=m_{0}$ ($E=m_{0}$) are at rest at infinity. They are coming from rest at infinity and
approaching the BH with different angular momenta $ L_{1}$ and $ L_{2}$ and collide at some radius $ r $. 
Our aim is to compute the energy in CM frame for this collision. We assume that two particles 1 and 2 are 
at same point with four momentum
\begin{equation}
p^{\mu}_{i} = m_{0} u^{\mu}_{i},
\end{equation}
where $p^{\mu}_{i}$ and $ u^{\mu}_{i} $ are the four momentum and four velocity of the $i-th$ particle ($i=1,2$). 
The sum of 2-momenta is given as
\begin{equation}
p^{\mu}_{tot} = p^{\mu}_{1} + p^{\mu}_{2},
\end{equation}
The CM energy of two-particles is given by
\begin{equation}\label{formula}
E_{CM}^2 = m_{1}^2+m_{2}^2- 2g_{\mu \nu} p_{1}^{\mu} p_{2}^{\nu},
\end{equation}
Clearly, $E_{CM}$ is scalar. The formula (\ref{formula}) is valid for both massless and massive particles, and 
independent of coordinate choice. In the CM energy case of equal mass $m_{1}= m_{2}=m_{0}$ the formula (\ref{formula}) reduces to
\begin{equation}\label{eqlm}
E_{CM} = m_{0} \sqrt{2} \sqrt{1-g_{\mu \nu} u^{\mu}_{(1)} u^{\nu}_{(2)}},
\end{equation}
Here we assume that the 2-particles coming from infinity with $E_{1}/m_{0}=E_{2}/m_{0}=1$, 
for simplicity. The CM energy of two equal mass particles in the rotating ABG spacetime is given by
\begin{widetext}
\begin{eqnarray}\label{ecm}
\frac{E_{CM}^2}{2 m_{0}^2} &=& \frac{1}{r^2(r^2 f(r) + a^2)} \Big[ a (f(r)-1)(L_{1} + L_{2})r^2 -a^2 (f(r)-3)r^2 - L_{1} L_{2} f(r)r^2 + (1 + f(r)) r^4
\nonumber \\
&-&\sqrt{-r^2\bigg((f(r)-1)a^2 -2a(f(r)-1)L_{1}-r^2+f(r)(L_{1}^2+r^2)\bigg)}
\nonumber \\
&&
\sqrt{-r^2\bigg((f(r)-1)a^2 -2a(f(r)-1)L_{2}-r^2+f(r)(L_{2}^2+r^2)\bigg)}\Big].
\end{eqnarray}
\end{widetext}
\subsection{Near horizon collision in extremal rotating ABG black hole}
Note that for an extremal BH, the two horizons always coincides and located at $r=r_{e}$,
where $r_{e}$ is the double root of Eq.~(\ref{eh}), i.e., $\Delta(r_{e})=r_{e}^2f(r_{e})+a^2=0$.
We now study the properties of CM energy Eq.~(\ref{ecm}) as radius $r \rightarrow r_{e}$ for the 
extremal rotating ABG spacetime. Firstly we find the range of angular momentum of particles with 
which particles can reach the horizon by solving Eq.~(\ref{lim}) numerically. The maximum/minimum 
angular momentum we are looking for extremal BH are listed in Table (\ref{table1}). We can see from 
the Table (\ref{table1}) that when charge $Q$ is increases, the extremal horizon $r_{e}$ is also increases.

We can see the behavior of CM energy for extremal BH from Fig.~\ref{fig8} and Fig.~\ref{fig10}. 
The Fig.~\ref{fig8} shows the variation of $E_{CM}$ vs $r$ for different values of $L_{1}$ and
$L_{2}$ with fixed values of spin parameter $a$ and charge $Q$.  We can see that $E_{CM}$ is infinite 
at the event horizon when $L_{1}$ is critical while for any other values of $L_{1}$, $E_{CM}$ is finite.
In Fig.~\ref{fig10} we show the variation of $E_{CM}$ vs $a$ at the horizon for different values of $L_{1}$
and $L_{2}$ with fixed value of charge parameter $Q$.
\begin{table}
\begin{center}
\caption{Numerical values of max/min angular momentum for extremal rotating ABG BH $M=1$, $m_{0}=1$, $E=1$.}\label{table1}
\resizebox{\linewidth}{!}{
\begin{tabular}{l l l l l l}
 \hline \hline
S.No.& $Q$   & $a$ & $r_{e}$  & $L_{min}$  & $L_{max}$ \\ 
\hline
1    & 0.1   & 0.980186499171  & 1.01388    &-4.80898    & 2.02893  \\  
2    & 0.2   & 0.922475716730  & 1.04536    &-4.75087    & 2.10710  \\ 
3    & 0.3   & 0.829553798960  & 1.07673    &-4.65290    & 2.22711  \\ 
4    & 0.4   & 0.700928370560  & 1.09464    &-4.50855    & 2.41043 \\
5    & 0.5   & 0.527525079404  & 1.08802    &-4.29774    & 2.71481  \\ 
 \hline \hline
\end{tabular}}
\end{center}
\end{table}
\begin{table}
\begin{center}
\caption{Numerical values of max/min angular momentum for non-extremal rotating ABG BH $M=1$, $m_{0}=1$, $E=1$.}\label{table2}
\begin{tabular}{l l l l l l}
 \hline \hline
S.No.& $Q$    & $a$   & $r_{+}$  & $L_{min}$  & $L_{max}$ \\ 
\hline
1    & 0.1    & 0.9   & 1.39848  & -4.75125     & 2.58660 \\  
2    & 0.2    & 0.8   & 1.49272  &-4.66002    & 2.77108 \\
3    & 0.3    & 0.7   & 1.50591  &-4.63034    & 2.52549  \\  
4    & 0.4    & 0.6   & 1.44629  & -4.42525    & 2.90270  \\ 
5    & 0.5    & 0.5   & 1.25836  &-4.27280     & 2.84983  \\ 
 \hline \hline
\end{tabular}
\end{center}
\end{table}
\subsection{Near horizon collision in non-extremal rotating ABG black hole}
A BH is called non-extremal when the outer and inner horizons are not coincide, i.e.
$r \neq r_{e}$. We find the range of angular momentum of particles by solving Eq.~(\ref{lim}) 
numerically. The maximum/minimum angular momentum we are looking for non-extremal BH are listed in Table (\ref{table2}). 

We can see the behavior of CM energy for non-extremal BH from Fig.~\ref{fig9} and 
Fig.~\ref{fig11}. In Fig.~\ref{fig9} we show the variation of $E_{CM}$ vs $r$ for different 
values of $L_{1}$ and $L_{2}$ with fixed values of spin parameter $a$ and charge $Q$. Here,
we can see that $E_{CM}$ is finite at the event horizon. In Fig.~\ref{fig11} we show the variation
of $E_{CM}$ vs $a$ at the outer horizon for different values of $L_{1}$ and $L_{2}$ with fixed values 
of charge $Q$. The variation of $E_{CM}$ vs $r$ for different values of $Q$ can be seen in panel (a) 
of Fig.~\ref{fig12}  and $E_{CM}$ vs $a$ in panel (a) of Fig.~\ref{fig13}. We have also plot the 
variation of $E_{CM}$ for Kerr BH in panel (b) of Fig.~\ref{fig12} and in panel (b) of Fig.~\ref{fig13}.

Eq.~(\ref{ecm}) confirms that the non-linear charge $Q$ indeed has influence on the CM energy. 
In the limit $a\rightarrow 0$, Eq.~(\ref{ecm}) reduces to expression for $E_{CM}$ of ABG BH,
\begin{widetext}
\begin{eqnarray}\label{ecm1}
\frac{E_{CM}^2}{2 m_{0}^2 }(a\rightarrow 0) &=& \Bigg[1+\frac{1}{f(r)}-\frac{L_{1} L_{2}}{r^2} -\frac{1}{f(r)}\sqrt{1-f(r)\left(1+\frac{L_{1}^2}{r^2}\right)} \sqrt{1-f(r)\left(1+\frac{L_{2}^2}{r^2}\right)}\Bigg].
\end{eqnarray}
\end{widetext}
which is exactly same as obtained in \cite{Pradhan:2014oaa}.
Further, $E_{CM}$ for the Kerr BH \cite{Banados:2009pr} can be obtained by taking limit $Q \rightarrow 0$ in 
Eq.~(\ref{ecm}), which reads
\begin{widetext}
\begin{eqnarray}\label{ecm2}
\frac{E_{CM}^2}{2 m_{0}^2 }(Q\rightarrow 0) &=& \frac{1}{r(r^2-2r+a^2)} \Big[2a^2 (1+r) - 2a (L_{1}+L_{2})- L_{1} L_{2}(-2+r) 
+2(-1+r)r^2
\nonumber \\
&-&\sqrt{2(a-L_{1})^2 - L_{1}^2r + 2r^2}\sqrt{2(a-L_{2})^2 - L_{2}^2r + 2r^2}\Big].
\end{eqnarray}
\end{widetext}
It turns out that Eq.~(\ref{ecm2}) the $ E_{CM}^2 $ blows up at the horizon in the extremal case \cite{Banados:2009pr}.
\section{Conclusion}
We are far from a reliable candidate for a quantum theory of gravity, hence recently more attention is given for  
phenomenological approaches  to somehow solve these singularity problem in classical general relativity and study possible 
implications. In this context, an important line of research is represented by the work on the regular BH solutions.   
Hence, we have investigated the collision of two particles
falling freely from rest at infinity in the background of a rotating regular ABG BH. It was demonstrated 
by BSW \cite{Banados:2009pr} that CM energy of two colliding particles near the horizon of an extremal 
Kerr BH can be arbitrarily high value under some restriction on the angular momentum of particles. 
The analysis of BSW when extended to rotating ABG BH to investigate the CM energy $E_{CM}$ for two particles, 
an unlimited $E_{CM}$ that constraints on the charge $Q$ and rotation parameter $a$. From the point of
view of BSW mechanism the CM energy for two colliding particles for extremal BH should be arbitrarily high. 
So that we have found critical value of the angular momentum for which the CM energy of colliding particles
is arbitrarily high near the horizon (extremal BH). We have also seen that when the angular momentum of the 
colliding particles is not critical (non-extremal BH) the CM energy is finite. For the non-extremal case, 
we find that the CM energy $E_{CM}$ increases with increase in the charge $Q$.  The range of the angular momentum 
$L$ and the spin parameter $a$ increases for the rotating ABG BH as compared with the Kerr-Newman BH. The CM energy $E_{CM}$
for the extremal rotating ABG BH case is infinite at $r_{e}=1.01388$ while for the extremal Kerr BH case the CM energy $E_{CM}$
is infinite at $r_{e}=1$.

Indeed, to obtain an arbitrary high $E_{CM}$, besides the conditions that the BH is an extremal BH, and one
of the particle has critical angular momentum, one still has restriction on the charge $Q$.  It may be pointed
out that horizon structure of the rotating ABG BHs is significantly complicated to analyze analytically and hence we turn
to numerically methods to analyze $E_{CM}$, the range of the angular momentum, with which particle can reach
horizon and fall into BH.

It turns out that if the angular momentum of a particle lies outside the range, it will not fall into the BH
and no collision will take place. We have also estimated the $E_{CM}$ for non-extremal rotating ABG BH case,
and found that the $E_{CM}$ is always finite and significantly affected by charge $Q$ and rotation parameter 
$a$. It will be of interest to extend the analysis discussed to other regular BH, which will be determine 
whether these properties of regular BHs are generic. These and related work are subject of forthcoming papers. 
\begin{acknowledgements}
M.A. acknowledges University Grant Commission, India for financial support through Maulana Azad National Fellowship 
scheme (File No.: F1-17.1/2012-13/MANF-2012-13-MUS-RAJ-8679).
\end{acknowledgements}

\end{document}